\documentclass[twocolumn,superscriptaddress,aps,preprintnumbers,amsmath,amssymb,sort&compress,nofootinbib
]{revtex4-1}
\usepackage{amsfonts,amsmath,amssymb,ascmac,bm,tensor}
\usepackage{comment}
\usepackage{ifpdf}
\usepackage{slashed}
\usepackage{color}
\usepackage[mathscr]{eucal}
\usepackage[utopia]{mathdesign}
\usepackage[utf8]{inputenc}
 
\usepackage{cancel}

\ifpdf
  \usepackage{graphicx}     %   usepackage without driver option
  \usepackage[bookmarksopen,colorlinks=true,linkcolor=bblue,citecolor=bblue,urlcolor=ppink]{hyperref}
\else     % For (p)LaTeX + dvipdfmx
  \usepackage[dvipdfmx]{graphicx}     %   usepackage with driver option
  \usepackage[dvipdfmx,bookmarksopen,colorlinks=true,linkcolor=bblue,citecolor=ppink,urlcolor=darkred]{hyperref}
\fi

\definecolor{red}{rgb}{1,0,0}
\definecolor{darkred}{rgb}{0.6,0,0}
\definecolor{darkgreen}{rgb}{0.992447,0.623778,0.034597}
\definecolor{ppink}{rgb}{1,0.4,0.4}
\definecolor{bblue}{rgb}{0.284602,0.317763,0.963947}
\newcommand{\1}{\mbox{1}\hspace{-0.25em}\mbox{l}}

\newcommand{\prn}[1]{\left( {#1} \right)}

\newcommand{\dd}{\mathrm{d}}
\newcommand{\Mpl}{M_{\rm Pl}}

\def\Mpl{M_{\rm Pl}}

\newcommand{\ee}{\mathrm{e}}

\newcommand{\calO}{\mathcal{O}}
\newcommand{\calP}{\mathcal{P}}

\makeatletter
\newcommand\footnoteref[1]{\protected@xdef\@thefnmark{\ref{#1}}\@footnotemark}
\makeatother

\begin{document}

%%%%%%%%%%%%%%%%%%%%%%%%%%%
%%%%%%%%%%% Title %%%%%%%%%%%
%%%%%%%%%%%%%%%%%%%%%%%%%%%

%%paper
\title{
$\mathcal{O}(10) M_\odot$ primordial black holes and string axion dark matter
}
\author{Keisuke Inomata}
\affiliation{ICRR, University of Tokyo, Kashiwa, 277-8582, Japan}
\affiliation{Kavli IPMU (WPI), UTIAS, University of Tokyo, Kashiwa, 277-8583, Japan}
\author{Masahiro Kawasaki}
\affiliation{ICRR, University of Tokyo, Kashiwa, 277-8582, Japan}
\affiliation{Kavli IPMU (WPI), UTIAS, University of Tokyo, Kashiwa, 277-8583, Japan}
\author{Kyohei Mukaida}
\affiliation{Kavli IPMU (WPI), UTIAS, University of Tokyo, Kashiwa, 277-8583, Japan}
\author{Yuichiro Tada}
\affiliation{ICRR, University of Tokyo, Kashiwa, 277-8582, Japan}
\affiliation{Kavli IPMU (WPI), UTIAS, University of Tokyo, Kashiwa, 277-8583, Japan}
\affiliation{IInstitut d’Astrophysique de Paris, UMR-7095 du CNRS, Universit\'e Pierre et Marie Curie, 98 bis bd Arago, 75014 Paris, France}
\affiliation{Sorbonne Universit\'es, Institut Lagrange de Paris, 98 bis bd Arago, 75014 Paris, France}
\author{Tsutomu T.~Yanagida}
\affiliation{Kavli IPMU (WPI), UTIAS, University of Tokyo, Kashiwa, 277-8583, Japan}
\affiliation{Hamamatsu Professor}

\begin{abstract}
\noindent
LIGO-Virgo collaboration has found black holes as heavy as $M \sim 30M_\odot$
through the detections of the gravitational waves emitted during their mergers.
Primordial black holes (PBHs) produced by inflation could be an origin of such BHs.
While it is tempting to presume that these PBHs constitute all Dark Matter (DM),
there exist a number of constraints for PBHs with $\mathcal{O} (10) M_\odot$ 
which contradict with the idea of PBHs as all DM.
Also, it is known that weakly interacting massive particle (WIMP) that is a common DM candidate is almost impossible to coexist with PBHs.
These observations motivate us to pursue another candidate of DM.
In this paper, we assume that the string axion solving the strong CP problem makes up all DM,
and discuss the coexistence of string axion DM and inflationary PBHs for LIGO events.
\end{abstract}

\date{\today}
\maketitle
\preprint{IPMU 17-0133}

%%%%%%%%%%%%%%%%%%%%%%%%%%%%%%%%%
%%%%%%%%%%% Introduction %%%%%%%%%%%
%%%%%%%%%%%%%%%%%%%%%%%%%%%%%%%%%
\section{Introduction}\label{sec: intro}

Axion appearing in String Theory~\cite{Conlon:2006tq,Svrcek:2006yi,Choi:2006za} 
is a plausible solution to the strong CP problem,
which may provide a platform 
to discuss why there exists an approximate global symmetry with an extreme accuracy
in the low-energy theory. 
The decay constant of axion is around the grand unified theory scale, $f_a \sim 10^{16}$\,GeV, in many string axion models.
Although a mild tuning of the misalignment angle is required not to overclose the Universe,
the coherently oscillating axion can account for the present dark matter (DM) abundance.\footnote{
	See Refs.~\cite{Kawasaki:1995vt,Kawasaki:2015pva} for discussion on the entropy production to avoid such a tuning.
}
Since axion is light, they may develop super-horizon fluctuations during inflation.
To avoid severe bounds on the isocurvature perturbation~\cite{Kawasaki:1997ct},
the Hubble parameter during inflation has to be low enough 
$H_\text{inf} \lesssim 
10^9 \, \text{GeV} \, \left( f_a /10^{16}\, \text{GeV} \right)^{0.405}.$

However, low-scale inflation has a serious drawback: 
namely an extreme fine-tuning of the initial condition is required (See \textit{e.g.},~\cite{Brandenberger:2016uzh}).
Interestingly, from the viewpoint of the string landscape,
such an initial condition can be selected dynamically.
In the string landscape, there are many different vacua with different vacuum energies,
and inflations may occur in each vacuum~\cite{Kachru:2003aw,Freivogel:2004rd,Susskind:2003kw,Bousso:2000xa}.
A preceding inflation takes place with a higher vacuum energy,
ends by tunneling or violating the slow-roll condition,
and the Universe falls into a vacuum with a smaller energy.
After successive such processes,
eventually, the low-scale inflation responsible for our observable patches of the Universe takes place.
To be concrete, let us consider hilltop inflation that is a prominent example of low-scale inflation.
If the inflaton for hilltop inflation acquires a positive Hubble-induced mass term, 
the inflaton can stay at the top of the flat potential during those pre-inflations.
As a result, the infamous initial condition problem is solved dynamically~\cite{Izawa:1997df},
which completes a consistent cosmological scenario for the string axion DM.

Recently, the LIGO-Virgo collaboration has announced the detections 
of three gravitational wave (GW) events, 
GW150914~\cite{Abbott:2016blz}, GW151226~\cite{Abbott:2016nmj}, and GW170104~\cite{Abbott:2017vtc}.
In addition to these three events, there is one GW candidate called LVT151012, whose 
signal-to-noise ratio is somewhat smaller than those of the other three GWs~\cite{Abbott:2016nmj}.
These GWs are generated by mergers of two black holes (BHs).
Their masses are $36.2\,^{+5.2}_{-3.8}$\,$M_\odot$ and $29.1\,^{+3.7}_{-4.4}$\,$M_\odot$ in GW150914, 
$14.2\,^{+8.3}_{-3.7}$\,$M_\odot$ and $7.5\,^{+2.3}_{-2.3}$\,$M_\odot$ in GW151226, 
$31.2\,^{+8.4}_{-6.0}$\,$M_\odot$ and $19.4\,^{+5.3}_{-5.9}$\,$M_\odot$ in GW170104,
and $23\,^{+18}_{-6}$\,$M_\odot$ and $13\,^{+4}_{-5}$\,$M_\odot$ in LVT151012.
One can see that the three BHs out of the eight BHs are as heavy as $30 M_\odot$.
However, in the usual metallicity ($Z \sim Z_\odot$), stellar origin BHs may not be $30M_\odot$ due to mass loss by stellar wind~\cite{TheLIGOScientific:2016htt,Belczynski:2009xy,Spera:2015vkd}.
Currently, researchers are actively searching for the origin of these BHs.

Primordial Black Hole (PBH) is one of the candidates of such BHs~\cite{Bird:2016dcv,Clesse:2016vqa,Sasaki:2016jop,Eroshenko:2016hmn,Carr:2016drx}.\footnote{
In a low-metallicity environment, stellar origin BHs with $30 M_\odot$ can be produced~\cite{TheLIGOScientific:2016htt,Belczynski:2009xy,Spera:2015vkd}.
This kind of BH is another candidate of the $30 M_\odot$ BHs~\cite{Kinugawa:2014zha}.
}
%\Blue{
It can be formed by the collapse of overdense Hubble patch in the early radiation-dominated 
universe~\cite{Hawking:1971ei,Carr:1974nx,Carr:1975qj}, which is completely different from the standard formation mechanism of stellar BHs
with supernovae. That is why PBH can have various masses such as $30M_\odot$. 
Also, it is clear that inflation can be an origin of such large density perturbations~\cite{GarciaBellido:1996qt,Kawasaki:1997ju,Yokoyama:1998pt}.
To realize PBHs however, the curvature perturbations on small scales 
should be strongly amplified roughly by $10^4$ times compared to those on the cosmic microwave background (CMB) scale,
though such a rapid change of the perturbation amplitude is generally difficult in the slow-roll single-field inflation~\cite{Motohashi:2017kbs}.
This is because the tilt of the power spectrum of the curvature perturbations are determined by the slow-roll parameters, which are small during slow-roll inflation.
On the other hand, if we have another inflation during the last 50--60 $e$-folds as supported by the viewpoint of the string landscape,
the second inflation can be free from the COBE normalization.
Therefore, although fine-tuning of the parameters is required as we show in Sec.~\ref{sec:newinf}, such an extreme modulation of perturbations can be 
achieved easily compared to ordinary single-field inflations~\cite{Kawasaki:1997ju,Kawasaki:1998vx,Frampton:2010sw,Kawasaki:2012kn,Kawasaki:2016ijp,Kawasaki:2016pql}.

Moreover, the axion DM and PBHs can coexist,
while the weakly interacting massive particle (WIMP) cannot~\cite{Lacki:2010zf,Bringmann:2011ut}.
Though the minimal setup might be the case where PBHs for the LIGO events account for
the present DM simultaneously, there are many constraints on this PBH mass range
which strongly disfavor PBHs as all DM.
Hence, we focus on the possibility that PBHs and other DM candidate coexist.
Since PBHs themselves and/or primordial perturbations large enough for abundant PBHs  
cause compact DM clumps~\cite{fillmore1984self,bertschinger1985self,Ricotti:2009bs,ludlow2010secondary,vogelsberger2009caustics}, 
annihilations of DM in the present Universe
can be dramatically enhanced, which makes the mixture of PBHs and WIMP DM difficult.
In this sense, axion is suitable because its interaction is feeble.

In this paper, we assume that $\mathcal{O} (10) M_\odot$ BHs detected by LIGO-Virgo collaboration are PBHs,
and construct a concrete scenario where the string axion DM and PBHs for the LIGO events coexist.
The organization of this paper is as follows.
In Sec.~\ref{sec:axion_isocurvature}, we review the string axion DM and the isocurvature perturbations produced by the fluctuations of an axion field.
We also discuss the properties of the inflation model compatible with the string axion DM.
In Sec.~\ref{sec:pbh}, we discuss PBHs for the LIGO events. 
The DM candidate in the presence of $\mathcal O(10)M_\odot$ PBHs is also discussed.
In Sec.~\ref{sec:newinf}, we construct the scenario consistent with the string axion DM.
We take the double inflation model as a concrete example and show the concrete parameter consistent with the observations.
Sec.~\ref{sec:conclusion_a} is devoted to the conclusion.

%%%%%%%%%%%%%%%%%%%%%%%%%%%%%%%%%%%%
\section{String axion DM and isocurvature perturbation}
\label{sec:axion_isocurvature}
%%%%%%%%%%%%%%%%%%%%%%%%%%%%%%%%%%%%

In this section, we summarize basic cosmological properties of the string axion.
Throughout this paper, we adopt the following normalization of the axion interaction:
\begin{align}
	\mathcal L_{a F \tilde F} 
	= \frac{g^2}{16 \pi^2 f_a}\, a\, \mathrm{Tr} \, F_{\mu\nu} \tilde F^{\mu\nu}.
\end{align}
For many string axion models, the axion decay constant tends to be large,
and the value is around the GUT scale~\cite{Conlon:2006tq,Svrcek:2006yi,Choi:2006za}:
\begin{align}
	f_a \sim 10^{16}\, \text{GeV}.
\end{align}
We take it as a fiducial value.

\paragraph*{\bf String axion as DM.}
Suppose that the moduli field (saxion) is stabilized throughout the evolution of our Universe.
Around the QCD phase transition, the string axion acquires a mass term via the QCD instanton,
and its potential becomes $\sim m_a^2 (T) a^2$ 
with $m_a (T)$ being the temperature dependent axion mass.
After this mass term exceeds the Hubble parameter, 
it starts to oscillate and behaves as cold dark matter (CDM).
In this case, the current energy density is determined by the initial amplitude,
namely the summation of the initial misalignment and the superhorizon fluctuations during inflation~\cite{Kawasaki:2013ae}:
\begin{align}
	\Omega_a h^2 \simeq
	10^4 \left( \frac{f_a}{10^{16}\,\text{GeV}} \right)^{1.19}
	\left( \theta^2 + \sigma_\theta^2 \right),
	\label{eq:omega_a}
\end{align}
where  $\theta$ is the misalignment angle
and $\sigma_\theta^2$ is the variance of the fluctuations given by\footnote{
	Precisely speaking, there is an additional logarithmic factor,
	$\ln(k_\text{max} L)$ with $k_\text{max}$ and $L$ being
	a comoving wavenumber entering the horizon at the end of inflation and the infrared cut off respectively.
	Though, parametrically, its value is $\mathcal O (10)$,
	the precise value is not so important for our purpose
	because we expect $\theta \gg \sigma_\theta$
	if the string axion dominates the present DM density,
	as we will see in the subsequent discussion.
}
\begin{align}
	\sigma_\theta^2 
	\simeq \left(  \frac{H_\text{inf}}{2 \pi f_a}  \right)^2.
\end{align}

The present DM density is $\Omega_\text{DM} h^2 \simeq 0.12$~\cite{Ade:2015xua}.
By requiring $\Omega_a \leq \Omega_\text{DM}$ and using Eq.~(\ref{eq:omega_a}),
we can set an upper bound on the misalignment angle~\cite{Kawasaki:2013ae,Kawasaki:2015pva},
\begin{align}
	\theta \lesssim 3.4 \times 10^{-3} \left( \frac{f_a}{10^{16}\,\text{GeV}} \right)^{-0.595},
\end{align}
and the Hubble parameter during inflation,
\begin{align}
	H_\text{inf} \lesssim 2 \times 10^{14} \, \text{GeV} 
	\left( \frac{f_a}{10^{16}\,\text{GeV}} \right)^{0.405}.
	\label{eq:axion_abun_iso}
\end{align}
As we show below, the constraint on the isocurvature perturbation
puts a more stringent bound on the Hubble parameter during inflation.

\paragraph*{\bf Isocurvature  perturbation.}

The axion condensate starts to oscillate when its mass exceeds the Hubble parameter.
Thus, for a homogeneous initial field value of axion,
its density perturbation is necessarily adiabatic.
However, since the mass of axion is light, 
the axion acquires fluctuations during inflation and 
the fluctuations yield differences in the initial condition for each Hubble patch at the onset of its oscillation.
As a result, the axion also develops the isocurvature perturbation.
The power spectrum of the axion isocurvature perturbation can be estimated as~\cite{Kawasaki:2013ae}
\begin{align}
	\mathcal P_\text{iso} \simeq 4\left( \frac{\Omega_a}{\Omega_\text{DM}}  \right)^2
	\frac{\left( H_\text{inf} / 2 \pi f_a \right)^2}{\theta^2 + \sigma_\theta^2}.
	\label{eq:axion_iso}
\end{align}
By analyzing the CMB spectrum, the Planck collaboration sets an upper bound on the isocurvature perturbation:
%%
%\Blue{
\begin{align}
	\beta_\text{iso} \equiv \frac{\mathcal P_\text{iso}}{\mathcal P_\text{ad}+\calP_\text{iso}}
	< 0.038
	\label{eq:iso}
\end{align}
%}
%%
at $95$\% CL for the scale of $k = 0.05$\,Mpc$^{-1}$~\cite{Ade:2015lrj}.
Here $\mathcal P_\text{ad} (\simeq 2.1 \times 10^{-9}) $ is the power spectrum of the adiabatic perturbation.
For the case where the string axion constitutes all the DM,
we can derive an upper bound on the Hubble parameter during inflation from Eqs.~(\ref{eq:axion_iso}) and (\ref{eq:iso}) as~\cite{Kawasaki:2015pva}
\begin{align}
	H_\text{inf} \lesssim 
	10^9 \, \text{GeV} \left( \frac{f_a}{10^{16}\, \text{GeV}} \right)^{0.405}.
	\label{eq:upper_bound_Hinf}
\end{align}
Note that, in the case where the string axion with $f_a \sim 10^{16}$\,GeV
makes up all the DM,
we expect $\theta \gg \sigma_\theta$, and
$\mathcal P_\text{iso} \simeq 4 (H_\text{inf}/2\pi f_a \theta)^2$.
This also means that the isocurvature constraint, Eq.~\eqref{eq:upper_bound_Hinf}, sets a severer bound than that of Eq.~\eqref{eq:axion_abun_iso}.

\paragraph*{\bf Hilltop inflation after pre-inflations.}
We have seen that the inflation scale has to be low ($H_\text{inf} \lesssim 10^{9}$\,GeV)
to avoid the severe constraints on the isocurvature perturbation for the string axion.
Hilltop (New) inflation~\cite{Boubekeur:2005zm,Linde:1981mu,Albrecht:1982wi} is one of the prominent examples of the low-scale inflation
 and thus we consider it in the following.

The major drawback of hilltop inflation is its initial condition problem.
If we assume that 
any inflationary phase does not precede hilltop inflation,
then there seems to be no reason to expect that the inflaton is smoothened over the horizon and its expectation value lies near a local minimum of its potential.
However, once we realize that there could be multiple inflations before the observed one, this problem can be solved dynamically.
Also, the picture of multiple pre-inflations may be natural in the context of the string landscape~\cite{Kachru:2003aw,Freivogel:2004rd,Susskind:2003kw,Bousso:2000xa}.

To make our discussion concrete, let us consider a pre-inflation right before the observed one.
An inflaton $\varphi$ which is responsible for the hilltop inflation may acquire a positive Hubble-induced mass term during the pre-inflation via $V_\text{pr} \varphi^2 / \Mpl^2$ with $V_\text{pr}$ being a potential for the pre-inflation.
If this is the case, $\varphi$ is stabilized at the local minimum throughout the pre-inflation,
and hence the initial condition is set dynamically. A concrete realization can be found in Ref.~\cite{Izawa:1997df}. There, it was assumed that a \textit{single} hilltop inflation after the pre-inflation explains the whole $e$-folds.
In general, there might be \textit{two (or more)} low-scale inflations that account for the required $e$-folds in total~\cite{Kawasaki:1997ju,Kawasaki:1998vx,Frampton:2010sw,Kawasaki:2012kn,Kawasaki:2016ijp,Kawasaki:2016pql}.
This consideration opens up a possibility to generate PBHs as we will see in the following sections.

%%%%%%%%%%%%%%%%%%%%%%%%%%%%%%%%%%%%
\section{Primordial Black Holes and the LIGO events}
\label{sec:pbh}
%%%%%%%%%%%%%%%%%%%%%%%%%%%%%%%%%%%%

In this section, we discuss a possibility to explain the LIGO GW events by mergers of PBHs
and the required property for the primordial density perturbation to have such PBHs.
We also discuss DM candidates which can coexist with PBHs.

\paragraph*{\bf LIGO GW events.}
The LIGO-Virgo collaboration has detected three GW events and one candidate from mergers of BH-binary.
Though the mass distribution is still unclear, those BHs have masses of $\mathcal O (10) M_\odot$.
They have also estimated the merger rate as $12$--$213$\,Gpc$^{-3}$\,yr$^{-1}$~\cite{Abbott:2017vtc}.
The question is an amount of PBHs required to explain these events.
This is addressed in Ref.~\cite{Sasaki:2016jop}; the fraction of PBHs has to be around 
$\mathcal O(10^{-3})$--$\mathcal O(10^{-2})$ to reproduce the merger rate.\footnote{
There are some papers that discuss the constraints on the $\mathcal O(10) M_\odot$ PBH abundance from the merger rate 
and the constraints are $\Omega_\text{PBH}/\Omega_\text{DM}< \mathcal O (10^{-2})$~\cite{Raidal:2017mfl,Ali-Haimoud:2017rtz}.
}
Unfortunately, the estimated event rate might suggest that those PBHs do not constitute all the DM.
However,
at the same time, the required amount of PBHs is still consistent with several observational constraints~\cite{Ali-Haimoud:2016mbv, Brandt:2016aco, Koushiappas:2017chw, Gaggero:2016dpq, Inoue:2017csr,monroy2014end,Carr:2017jsz,Kuhnel:2017pwq,Green:2016xgy,Inomata:2017okj}.

\paragraph*{\bf PBH formation.}
Here we briefly review the formation of PBHs by the primordial density perturbation
and discuss the suitable property of the primordial density perturbation
to account for the LIGO events.

Basically, PBHs are formed by the gravitational collapse of an over-dense region.
According to simple analysis by Carr~\cite{Carr:1975qj}, if a density contrast is larger than $\delta > 1/3$ at its horizon reentry,
 the over-dense region overcomes the radiation pressure and PBHs are formed.
We use this threshold value $\delta_c = 1/3$ as a fiducial value in the following.
The PBH mass is roughly estimated as the horizon mass at the horizon reentry of the perturbations which are larger than the threshold.
We can estimate the relation between the scale of the perturbations and PBH mass as
\begin{align}
	M(k)  
	&= \left. \gamma \rho \frac{4 \pi H^{-3}}{3} \right|_{k = aH}
	\simeq\frac{\gamma M_\text{eq}}{\sqrt{2}} 
	\prn{ \frac{g_{\ast,\text{eq}}}{g_\ast} }^\frac{1}{6}
	\prn{ \frac{k_\text{eq}}{k} }^2  \nonumber \\[.5em]
	&\simeq M_\odot\left(\frac{\gamma}{0.2}\right)\left(\frac{g_*}{10.75}\right)^{-\frac{1}{6}}\left(\frac{k}{1.9\times10^6\,\mathrm{Mpc}^{-1}}\right)^{-2} 
	\label{eq:pbhmass in k} \\[.5em]
	&\simeq M_\odot
    \left(
    \frac{\gamma}{0.2}
    \right)
    \left(
    \frac{g_\ast}{10.75}
    \right)^{- \frac{1}{6}}
    \left(
    \frac{f}{2.9 \times 10^{-9} \,\textrm{Hz
   }}
    \right)^{-2},
    	\label{eq:pbhmass}
\end{align}
where $\gamma$ is a numerical factor which depends on the properties of gravitational collapse 
and $M_\text{eq}$ is the horizon mass at matter-radiation equality time 
and $g_*$ and $g_{*,\text{eq}}$ are the degrees of freedom at PBH formation and matter-radiation equality time,
 where $g_*\simeq 10.75$ at $\mathcal O(10) M_\odot$ PBH production.
We adopt the simple analysis value $\gamma = (1/3)^{3/2}\simeq 0.2$ as a fiducial value~\cite{Carr:1975qj}.

Assuming that the density perturbations follow the Gaussian distribution, 
the production rate of PBHs with $M_\text{PBH} = M$ can be written as 
\begin{align}
	\beta (M) =
	\int_{\delta_c}
	\frac{\dd \delta}{\sqrt{2 \pi \sigma^2 (M)}} \, e^{- \frac{\delta^2}{2 \sigma^2 (M)}}
	\simeq 
	\frac{1}{\sqrt{2 \pi}} \frac{1}{\delta_c / \sigma (M)} \, e^{- \frac{\delta_c^2}{2 \sigma^2 (M)}}.
	\label{eq:beta}
\end{align}
$\sigma^2(M)$ is the variance of density contrast with the smoothing scale $M(k)$
which is defined as~\cite{Young:2014ana}
\begin{align}
	\sigma^2 (M (k))
	= \int \dd \ln q W^2 (q k^{-1}) \frac{16}{81} \prn{q k^{-1}}^4 \mathcal P_\zeta (q),
	\label{eq:sigma}
\end{align}
where $\mathcal P_\zeta(k)$ is the power spectrum of the primordial curvature perturbation.
$W(x)$ is a window function and throughout this paper we use the Gaussian window function $W(x) = \ee^{-x^2/2}$.

Once PBHs are formed, the PBH energy density behaves as non-relativistic matter and decays as $\rho_\text{PBH} \propto a^{-3}$.
On the other hand, the background energy density behaves as radiation $\rho \propto a^{-4}$ until the matter-radiation equality time.
Therefore the ratio of the PBH energy density to the background energy density grows as $\rho_\text{PBH}/\rho \propto a$.
Using the PBH production rate $\beta(M)$, we can derive PBH-DM ratio over logarithmic mass interval $\dd \, \text{ln} M$ as
\begin{align}
	\frac{\Omega_\text{PBH} (M)}{\Omega_\text{DM}}
	&\simeq  \left. \frac{\rho_\text{PBH}(M)}{\rho_m} \right|_\text{eq} \frac{\Omega_{m}}{\Omega_\text{DM}}
	= \prn{\frac{T_M}{T_\text{eq}} \frac{\Omega_{m}}{\Omega_\text{DM}}} \gamma \beta (M) \\
	& \simeq
	\!\left(\! \frac{\beta (M)}{1.84 \times 10^{-8}} \!\right)
	\!\left(\! \frac{\gamma}{0.2} \!\right)^\frac{3}{2}\!
	\!\left(\! \frac{10.75}{g_{\ast} (T_M)} \!\right)^\frac{1}{4}\!
	\!\left(\! \frac{0.12}{\Omega_\text{DM}h^2} \!\right)
	\!\left(\! \frac{M}{M_\odot} \!\right)^{-\frac{1}{2}} \hspace{-7pt}, 
	\label{eq:frac}
\end{align}
where we define $\Omega_\text{PBH}(M) \equiv \dd \Omega_\text{PBH}/\dd \, \text{ln} M$ and $\rho_\text{PBH}(M) \equiv \dd \rho_\text{PBH}/\dd \, \text{ln} M$.
$\Omega_m$ and $\Omega_\text{DM}$ are the energy density parameter of total matter (baryon + CDM) and CDM.
$T_M$ and $T_\text{eq}$ are the temperatures of the Universe at the horizon reentry of the perturbations and the matter-radiation equality time.
We adopt $\Omega_\text{DM} h^2 = 0.12$~\cite{Ade:2015xua}.
The total PBH-DM ratio is given by
\begin{align}
	\Omega_{\text{PBH,tot}} = \int \dd \ln M\, \Omega_\text{PBH} (M).
\end{align}

To sum up, we need a large curvature perturbation, 
$\mathcal P_\zeta (k) \sim 10^{-2}$,
at a scale of $k \sim 10^{6}$\,Mpc$^{-1}$
to reproduce the LIGO events by mergers of PBHs.
Since the curvature perturbation at the large scale, $k \lesssim 1$\,Mpc$^{-1}$,
should be as small as $\sim 10^{-9}$ not to conflict with the CMB observation,
we need to break the (almost) scale-invariance of the spectrum.
In addition, 
there are other constraints on such a large curvature perturbation at the scale, $k \sim 10^{6}$\,Mpc$^{-1}$, as pointed out in Ref.~\cite{Kohri:2014lza,Saito:2008jc,Saito:2009jt}; (i) CMB spectrum distortion and (ii) induced GWs by the second order effect.
Basically, the peak of the power spectrum of the curvature perturbations should be sufficiently sharp\footnote{
One can avoid such constraints also by an enhanced non-Gaussianity~\cite{Nakama:2016gzw,Nakama:2016kfq}.
}
 to avoid them
and one must explicitly check that the model is well consistent with those constraints to investigate a successful scenario.

\paragraph*{\bf DM candidates in the presence of PBHs.}
Throughout this paper, we focus on the string axion in a low-scale inflation scenario.
Thus, our primary DM candidate is axion.
Nevertheless, it is instructive to compare the axion DM with other DM candidates
in the presence of PBHs.

The fascinating possibility would be the case where PBHs for the LIGO events can constitute the whole DM simultaneously.
Unfortunately, it seems to be difficult to achieve full DM because of many observational constraints at this mass range~\cite{Ali-Haimoud:2016mbv, Brandt:2016aco, Koushiappas:2017chw, Gaggero:2016dpq, Inoue:2017csr,monroy2014end,Carr:2017jsz,Kuhnel:2017pwq,Green:2016xgy,Inomata:2017okj}.
Hence, we need to consider other DM candidates.

The WIMP is a major particle DM candidate.
However, it is known that WIMP is difficult to coexist with PBHs.
This is because a large density perturbation promotes the formation of DM halos 
with a steep profile, which is called Ultra-Compact Mini-Halos (UCMHs)~\cite{fillmore1984self,bertschinger1985self,
Ricotti:2009bs,ludlow2010secondary,vogelsberger2009caustics}.
As a result, the DM density at the center of the UCMH becomes so dense that
DM annihilations at the present Universe are dramatically enhanced.
Thus, we can constrain such scenarios by the observation of gamma-rays from DM annihilations.
The upper bound on the annihilation cross section is known to be
several orders of magnitude smaller than its typical value for WIMP, 
$\langle \sigma v \rangle  \sim 10^{-26}$\,cm$^{3}$s$^{-1}$~\cite{Bringmann:2011ut}.
Moreover, even if PBHs are not produced by large primordial perturbations,
the existence of PBHs itself leads to the formation of UCMHs with PBHs as their cores. Therefore the PBH abundance
is constrained as $\Omega_\text{PBH}\lesssim10^{-4}$ (for $m_\text{DM}\sim100\,\text{GeV}$) for a vast range in PBH mass
with the thermal relic DM~\cite{Lacki:2010zf}.
These facts indicate the difficulty of the coexistence of PBHs and WIMP DMs.

One important constraint on the coexistence of the string axion and BHs
comes from the superradiance effect~\cite{1971JETPL..14..180Z,Misner:1972kx,Starobinsky:1973aij}.
We briefly mention it here.
If a Compton wavelength of axion is close to the size of a rotating BH,
then such a BH immediately loses its spin by filling bounded Bohr orbits 
with a huge number of axions due to the bosonic nature of axions~\cite{Zouros:1979iw,Detweiler:1980uk,Gaina:1988nf}.
As a result, observations of rotating BHs can constrain axion models~\cite{Arvanitaki:2014wva,Arvanitaki:2016qwi}.
Parametrically, we expect that a given rotating BH mass can constrain
$m_a \sim 1/r_g = 4 \pi\Mpl^2 / M_\text{BH} \sim 10^{-11}\,\text{eV}\, (M_\odot / M_\text{BH})$,
or equivalently $f_a \sim 10^{18} \,\text{GeV}\, (M_\text{BH}/M_\odot)$.
One can see that ordinary stellar BHs with $M_\text{BH} \sim M_\odot$ 
cannot probe the axion decay constant much below the Planck scale.
Hence, $f_a \sim 10^{16}$\,GeV in our scenario is still viable.

%%%%%%%%%%%%%%%%%%%%%%%%%%%%%%%%%%%%%%%%%%%%
\section{Double inflation after pre-inflations}
\label{sec:newinf}
%%%%%%%%%%%%%%%%%%%%%%%%%%%%%%%%%%%%%%%%%%%%

As mentioned at the end of Sec.~\ref{sec:axion_isocurvature},
if we have two (or more) inflations which explain the required $e$-folds,
the primordial curvature perturbation can be large enough for a sizable amount of PBHs by breaking the scale-invariance
and generate PBHs.
In this paper, we consider the double inflation model~\cite{Kawasaki:1997ju} as a concrete example.\footnote{
While our model causes double inflation with two fields, some inflation models cause double inflation with single field~\cite{Kannike:2017bxn,Ballesteros:2017fsr}.
In~\cite{Kannike:2017bxn,Ballesteros:2017fsr}, they discuss PBHs with their mass around $\mathcal O(10^{20}) \text{g}$.
}
In this section, first, we search for the desired parameter regions semi-analytically, taking account of two conditions.
One is that the resonance enhancement of perturbations are small enough for a linear analysis.
The other one is that the first inflation should correctly reproduce the result of the Planck observations on the CMB scale.
After showing the desired parameter regions, we will show the fully numerical result with one desired parameter set in the last subsection.

\paragraph*{\bf Inflation Scenario.}
The double inflation model has two stages of inflation. 
During the intermediate phase between the first inflation and the second inflation, this model can generate the large perturbations which produce PBHs.
Note that the large-scale perturbations observed by the Planck satellite are generated by the first inflation and the small-scale perturbations are generated by the second inflation.

As discussed in Sec.~\ref{sec:axion_isocurvature},
the string axion DM requires the low energy scale of the first inflation as
\begin{align}
	H_\text{inf,first} \lesssim 10^{9} \,\text{GeV},
	\label{eq:first_inf_energy_cons}
\end{align}
because the isocurvature perturbations generated by the first inflation are severely constrained by CMB observations.
In the following, we assume that the first and second inflations are low-scale hilltop inflations to satisfy Eq.~(\ref{eq:first_inf_energy_cons}).

We take the following inflaton potentials:
\begin{align}
	\label{eq:inf_potential_tot}
	V(\varphi,\chi) 
	&= V_1 (\varphi) + V_2 (\chi) + V_\text{stb} (\varphi,\chi) + C, \\
	\label{eq:inf_potential_new1}
	V_1 (\varphi) 
	&= v_1^4\left(  -\frac{\varphi}{\varphi_l} %- 2\sqrt{2}c_1v_1^2 \varphi
	- \frac{\varphi^2}{\varphi_q^2}
	+ \left( 1 - \frac{\varphi^{n_1}}{\varphi_\text{min} ^{n_1}} \right)^2  \right), \\
	\label{eq:inf_potential_new2}
	V_2 (\chi) 
	&= v_2^4 \left(- \frac{\chi}{\chi_l} %- 2\sqrt{2}c_1v_1^2 \varphi
	-  \frac{\chi^2}{\chi_q^2}
	+ \left( 1 - \frac{\chi^{n_2}}{\chi_\text{min} ^{n_2}} \right)^2 \right), \\
	\label{eq:inf_potential_stb}
	V_\text{stb} (\varphi,\chi) 
	&= \frac{1}{2}c_\text{pot} V_1 (\varphi) \chi^2,
\end{align}
where $M_\text{Pl}$ is set to be unity and $\varphi$ is the inflaton of the first inflation and $\chi$ is the inflaton of the second inflation. 
$C$ is a compensate term which makes $V(\varphi_\text{min}, \chi_\text{min}) = 0$.\footnote{
Strictly speaking, the true potential minimum is not $V(\varphi_\text{min}, \chi_\text{min})$.
However, the deviation of the true potential minimum from $V(\varphi_\text{min}, \chi_\text{min})$ is so small 
and we have checked that the deviation does not affect our results.
}
We take $n_1, n_2 \geq3$.\footnote{
If we take $n_1, n_2=2$, we must take the value of inflaton at the potential minimum as $\varphi_\text{min}>M_\text{Pl}$ or $\chi_\text{min}>M_\text{Pl}$
 to realize slow-roll inflation.
If an inflaton becomes larger than Planck scale, the inflaton potential can be modified by the other Planck suppressed operators we don't write explicitly
 and it can be difficult to follow the dynamics of the inflaton.
 Therefore we restrict our discussion to the case of $n_1,n_2\geq 3$.
}
Generally speaking, we also expect Planck-suppressed order correction to kinetic terms,
\begin{align}
	\mathcal{L}_\text{kin} = -\frac{1}{2} \left( 1- \frac{c_\text{kin}}{2} \chi^2 \right) \partial_\mu \varphi \partial^\mu \varphi 
	 - \frac{1}{2}  \partial_\mu \chi \partial^\mu \chi + \cdots.
	\label{eq:inf_kinetic}
\end{align}
The expected inflation dynamics is described as follows.
First, the first inflation takes place while $\varphi$ slowly rolls down the potential and the energy scale of the first inflation is $\sim v_1^4$.
During the first inflation, the other inflaton $\chi$ is stabilized at $\chi \simeq v_2^4/(c_\text{pot} v_1^4 \chi_l)$ because of the stabilization term, $\frac{1}{2} c_\text{pot} V_1(\varphi) \chi^2$.
After a while, the slow-roll of $\varphi$ ends and $\varphi$ starts to oscillate around the potential minimum, $\varphi_\text{min}$.
During this time, $\varphi$ possibly passes many times through the tachyonic region, where the mass of $\varphi$ becomes tachyonic, 
and non-adiabatic region, where the adiabatic condition is violated,
and the perturbations of $\varphi$ can grow so large that the perturbations become non-linear.
We discuss this issue again in the next subsection and App.~\ref{sec:resonance}. 
After the amplitude of the oscillation becomes small due to the Hubble friction, the energy density deposited in $\varphi$ becomes smaller than the second inflation energy scale $\sim v_2^4$ and then the second inflation starts.
By then, the stabilization of $\chi$ becomes so weak that $\chi$ can roll down the potential.
After the slow-roll of $\chi$ ends, $\chi$ oscillates around the minimum and decays to other particles at least via Planck-suppressed operators.
Around minimum, the mass of $\chi$ is given by
\begin{align}
	m_\chi = \sqrt{2} n_2 \frac{v_2^2}{\chi_\text{min}}.
	\label{eq:chi_mass}
\end{align}
Then the corresponding reheating temperature can be evaluated as
\begin{align}
	T_\text{R} 
	\simeq 0.1 m_\chi^{3/2} 
	\simeq 0.2 n_2^{3/2} \frac{v_2^3}{\chi_\text{min}^{3/2}},
	\label{eq:reheating_temp}
\end{align}
if the reheating is achieved by some Planck-suppressed operators.
We use this value in the following analysis.

\begin{figure}
	\centering
	\includegraphics[width=.40\textwidth]{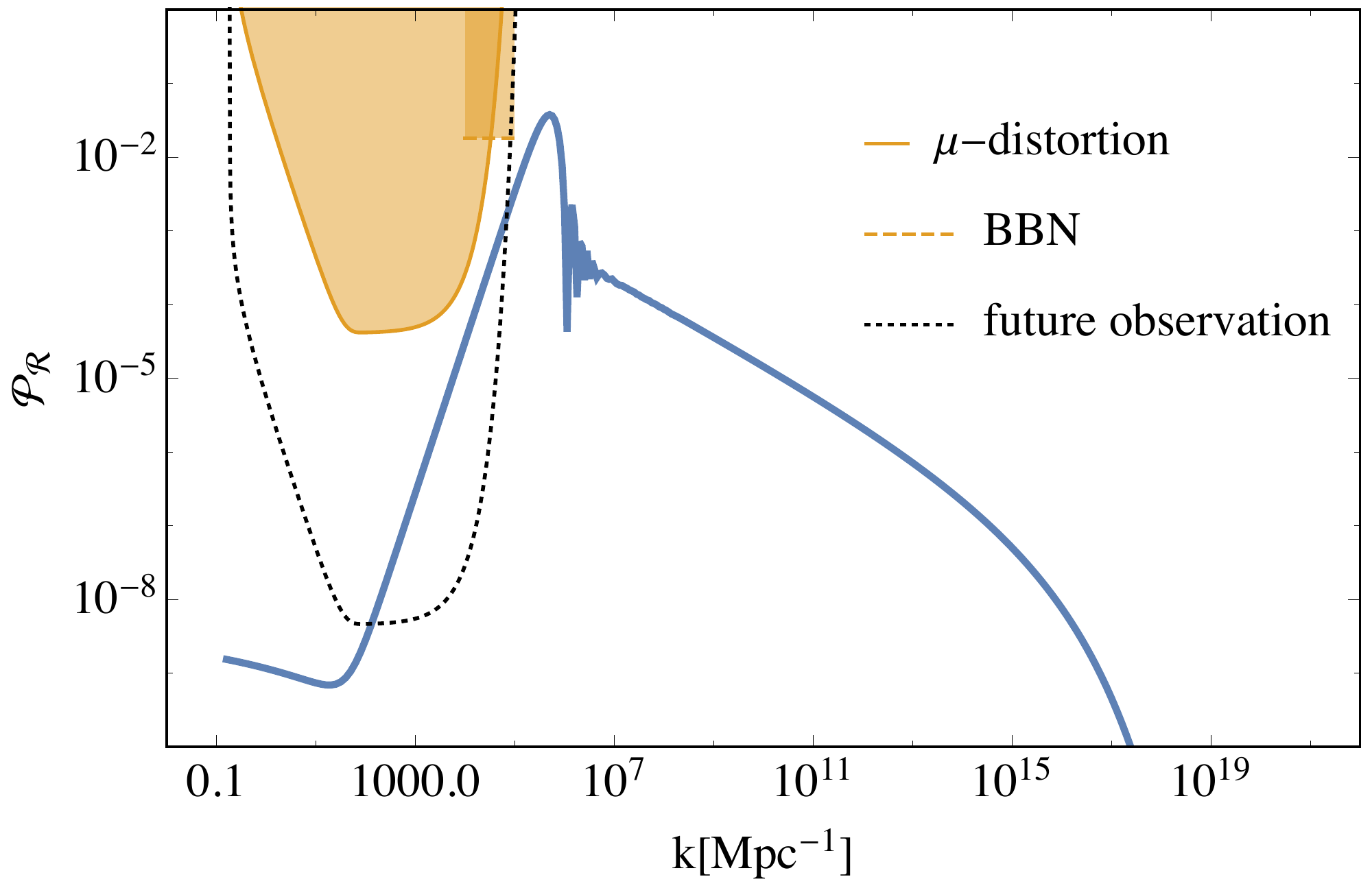}
	\caption{\small
	The scalar power spectrum for parameters given in Eq.~(\ref{eq:final_para}).
	Orange shaded regions are excluded by 
	the current constraint on $\mu$-distortion, $|\mu|<9\times10^{-5}$~\cite{Fixsen:1996nj}
	and the effect on n-p ratio during big-bang nucleosynthesis~\cite{Inomata:2016uip}.\footnote{
There are other constraints on the power spectrum of curvature perturbations around $10^{4}$--$10^{5}\, \text{Mpc}^{-1}$~\cite{Jeong:2014gna,Nakama:2014vla}
and their constraints are similar to~\cite{Inomata:2016uip}.
}
	The black dotted line represents a future constraint by $\mu$-distortion with the PIXIE~\cite{Kogut:2011xw}, 
	$|\mu|<10^{-8}$.
	}
	\label{fig:curvature_n3}
\end{figure}

Let us estimate the $e$-folds which corresponds to the peak scale of curvature perturbations. 
We define the pivot scale as $k_\text{pivot}=0.05$\,Mpc$^{-1}$ and the peak scale as $k_\text{peak} = 5 \times 10^5$\,Mpc$^{-1}$ (see Fig.~\ref{fig:curvature_n3}).
Then the difference between the corresponding $e$-folds is calculated as
\begin{align}
	N_\text{pivot} - N_\text{peak} 
	&= \text{log}\left( \frac{k_\text{peak}}{k_\text{pivot}} \right) \nonumber \\
	&\simeq 16.
	\label{eq:peak_e_fold}
\end{align}
Note that there is a difference between the scale at the first inflation end and the peak scale as discussed in 
App.~\ref{sec:multihorizon}. 
The difference between $N_\text{peak}$ and $N_\text{first inf, end}$ ($=$ $e$-folds at the end of the first inflation) is given by
\begin{align}
	\frac{H_\text{inf,first}}{H_\text{inf,second}} \simeq \ee^{3(N_\text{peak} - N_\text{first inf, end}) },
	\label{eq:end_peak_e_fold}
\end{align}
where $H_\text{inf,first}$ and $H_\text{inf,second}$ are the energy scales of the first inflation and the second inflation.
Here we define $N_\text{first,pivot}$ as $N_\text{first,pivot} \equiv N_\text{pivot} - N_\text{first inf, end}$.
If we take $H_\text{inf,first}/H_\text{inf,second} = 400$, 
we derive $N_\text{peak} - N_\text{first inf, end} \simeq 2$ and $N_\text{first,pivot} = 18$.
In the following, we use $H_\text{inf,first}/H_\text{inf,second} = 400$ and  $N_\text{first,pivot} = 18$ as fiducial values.

\begin{figure*}
  \begin{minipage}[b]{0.49\linewidth}
    \centering
    \includegraphics[keepaspectratio, scale=0.5]{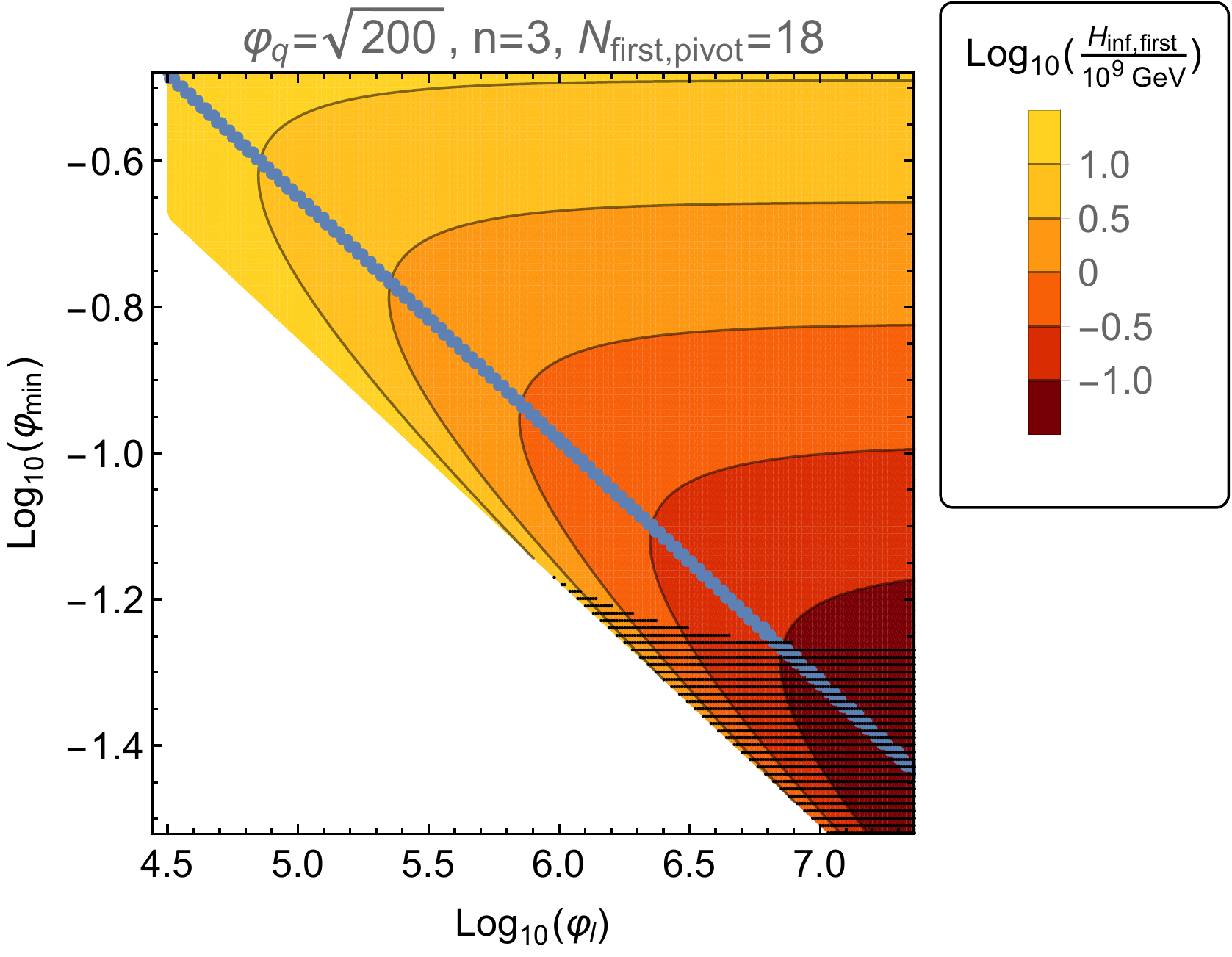}
  \end{minipage}
  \begin{minipage}[b]{0.49\linewidth}
    \centering
    \includegraphics[keepaspectratio, scale=0.5]{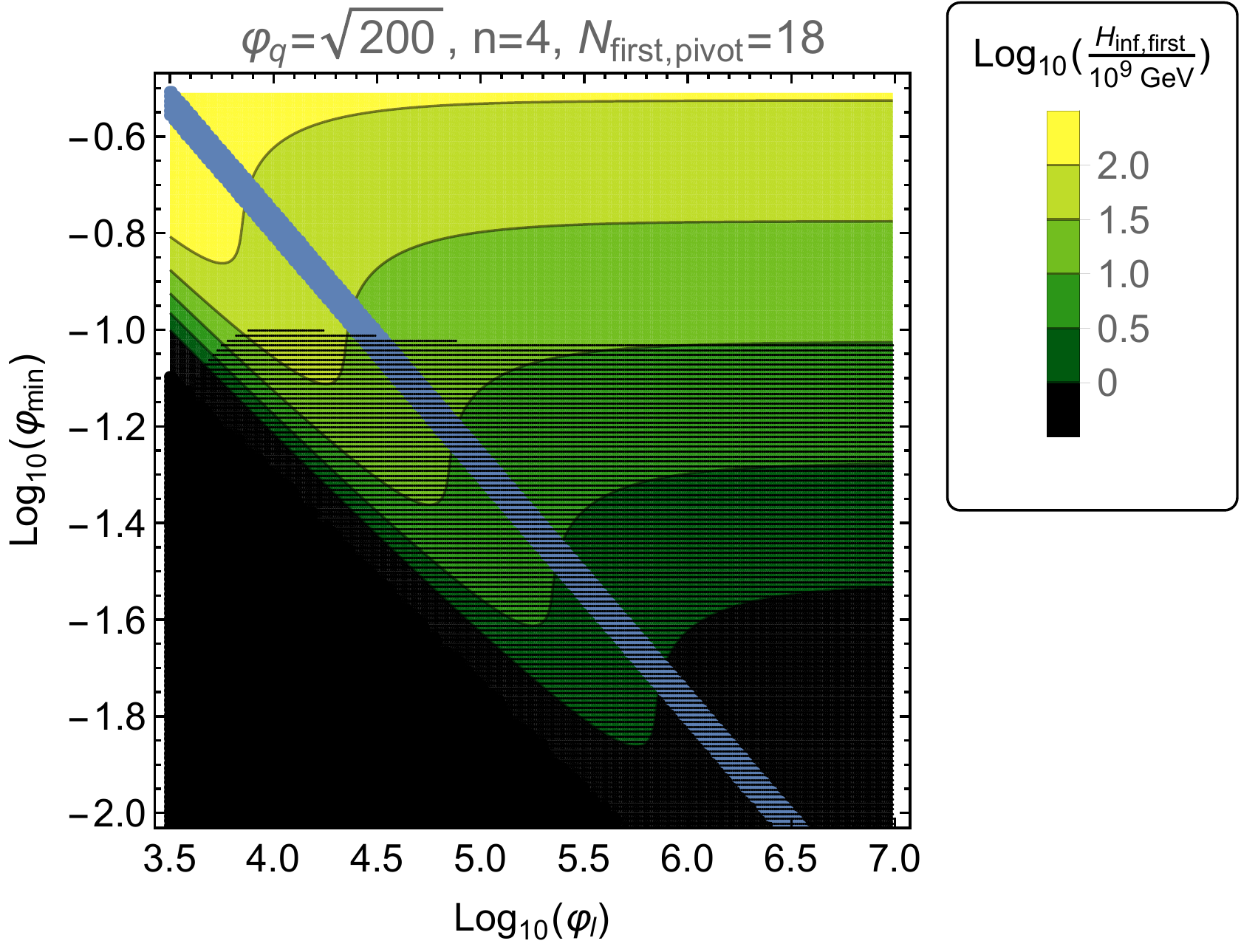}
  \end{minipage}
  \caption{\small
	The blue shaded regions show the parameter regions where both $n_s$ and $\dd \, n_s/\dd\, \text{ln}k$ are consistent with observations. 
	In each point of this figure, we calculate the value of $H_\text{inf,first}$ with which $A_s$ is consistent with the observation.
	The lower left white region in the left figure shows the parameter region of $N_\text{first, pivot}<18$.
	The light black shaded regions show the parameter regions in which perturbations become non-linear due to the resonances 
	(see also App.~\ref{sec:resonance} for details).
	We take $\varphi_q = \sqrt{200}$ and $N_\text{first,pivot}=18$ as a fiducial value and we take $n=3$ in the left figure and $n=4$ in the right figure.
	}	
	\label{fig:allowed_region}
\end{figure*}

 %%%%%%%%%%%%%%%%%%%%%%%%%%%%
\paragraph*{\bf Resonance.}
 %%%%%%%%%%%%%%%%%%%%%%%%%%%%
Some inflation models have the preheating phases with some resonances, which follow the inflation phases.
In general, hilltop inflation has a preheating phase driven by the tachyonic mass and non-adiabaticity after its inflation.
Throughout this paper, we focus on the preheating phase which occurs after the first inflation 
because we must follow the evolution of the perturbation which exits the horizon during the second inflation.
Roughly speaking, the resonance becomes stronger for smaller $\varphi_\text{min}$ and larger $H_\text{inf,first}$
because the ratio of the oscillation timescale to the Hubble friction timescale ($H_\text{inf,first}/m_\varphi = \sqrt{3/2}\varphi_\text{min}/n$) becomes smaller for smaller $\varphi_\text{min}$
and the seed perturbation $H_\text{inf,first}/2\pi$ becomes larger for higher $H_\text{inf,first}$. In Fig.~\ref{fig:allowed_region},
we show the parameter regions where the perturbations become non-linear, which is estimated by numerical calculations. 
We discuss the behavior of the resonances further in  App.~\ref{sec:resonance}.
If the resonances are strong and the perturbations become non-linear, it becomes complicated to calculate the evolution of perturbations 
because the evolution depends on the higher order perturbations and back-reactions of perturbations to the homogeneous backgrounds become non-negligible.
In the double inflation model we consider here,
there are the interaction terms between inflatons $\varphi$ and $\chi$, 
such as $\frac{1}{2} c_\text{pot} V_1(\varphi) \chi^2$ and $\frac{1}{4} c_\text{kin} \chi^2 \partial_\mu \varphi \partial^\mu \varphi$.
Hence, we must follow the dynamics of the inflaton $\varphi$ during its oscillation phase to calculate the perturbations.
In this paper, to avoid the complexities, we search for the parameters in which the perturbations remain linear until horizon exit.
The study of the non-linear regime is left for future works.

%%%%%%%%%%%%%%%%%%%%%%%%%%%%
\paragraph*{\bf CMB constraints.}
%%%%%%%%%%%%%%%%%%%%%%%%%%%%
Now, we return to the double inflation model and discuss the constraints on the large-scale perturbations.
From the Planck results, the cosmological parameters are determined in $95\%$ CL as ~\cite{Ade:2015xua}:
\begin{align}
	\label{eq:cosmo_par_A}	
	10^9 A_s &= 2.142 \pm 0.098,\\		
	\label{eq:cosmo_par_ns}	
	n_s &= 0.9667\pm0.0080,\\
	\label{eq:cosmo_par_run}	
	\frac{\dd \, n_s}{\dd\, \text{ln}k} &= -0.0085\pm 0.0152, \\
	\label{eq:cosmo_par_r}
	r_{0.002} &< 0.113,
\end{align}
where $A_s$ is the amplitude of the adiabatic scalar power spectrum, $n_s$ is the scalar spectral index, 
$\dd n_s/ \dd \text{ln} k$ is the running of scalar spectral index and $r_{0.002}$ is the tensor-to-scalar ratio.

Since the first inflation generates the curvature perturbations on the pivot scale, we must check whether the first inflation can be consistent with Eqs. (\ref{eq:cosmo_par_A})--(\ref{eq:cosmo_par_r}).
Thus we focus on the first inflation whose potential is given by Eq.~(\ref{eq:inf_potential_new1}).\footnote{
The relation between cosmological parameters and the parameters of this inflation potential has also been studied
in~\cite{Takahashi:2013cxa,Harigaya:2013pla}.}

The slow-roll parameters, $\varepsilon, \eta$ and  $\zeta$, are defined as
\begin{align}
	\label{eq:varepsilon_def}	
	\varepsilon &\equiv \frac{1}{2} \left( \frac{V_1'}{V_1} \right)^2, \\
	\label{eq:eta_def}	
	\eta &\equiv \frac{V_1''}{V_1}, \\
	\label{eq:zeta_def}	
	\zeta &\equiv \frac{V_1'V_1'''}{V_1^2},
\end{align}
where a prime denotes a derivative with respect to $\varphi$.
The cosmological parameters can be written with the slow-roll parameters as 
\begin{align}
	\label{eq:slow_par_A}	
	A_s & \simeq \frac{H_\text{inf,first}^2}{8\pi^2 \varepsilon},\\		
	\label{eq:slow_par_ns}	
	n_s & \simeq 1-6\varepsilon -2\eta,\\
	\label{eq:slow_par_run}	
	\frac{\dd \, n_s}{\dd\, \text{ln}k} &\simeq -24\varepsilon^2 + 16\varepsilon \eta -2 \zeta,\\
	\label{eq:slow_par_r}
	r_{0.002} &\simeq 16\varepsilon.
\end{align}
We can calculate the $e$-folds from the end of the first inflation with the following equation:
\begin{align}
	N(\varphi) = \int^\varphi_{\varphi_e} \frac{V}{V'} \dd \varphi,
	\label{eq:e_fold_formula}
\end{align}
where $\varphi_e = \left( \varphi_\text{min}^n/(2n_1(n_1-1)) \right)^{\frac{1}{n_1-2}}$.
According to the discussion in the previous subsection about resonance, 
we define $\varphi_\text{first,pivot}$ as the value satisfying $N(\varphi_\text{first,pivot})=18$.
If we determine $\varphi_l, \varphi_q$ and $\varphi_\text{min}$, we can calculate $\varphi_\text{first,pivot}$ numerically.
Then we can substitute it into Eqs.~(\ref{eq:varepsilon_def})--(\ref{eq:slow_par_r}) and derive $A_s, n_s, \dd \, n_s/\dd\, \text{ln}k$ and $r_{0.002}$.
Since the relation $\varepsilon \ll \eta$ is generally satisfied in hilltop inflation models,
Eq.~(\ref{eq:cosmo_par_r}) is always satisfied in our concrete inflation model
and we do not consider Eq.~(\ref{eq:cosmo_par_r}) explicitly in the following.
Here, let us stress that $n_s$ and $\dd n_s/\dd\, \text{ln} k$ are independent of $H_\text{inf,first}$.
This is because the slow-roll parameters, $\varepsilon, \, \eta$ and $\zeta$, are independent of $H_\text{inf,first}$ as
\begin{align}
	\varepsilon &\simeq \frac{1}{2} \left( -\frac{1}{\varphi_l} -2 \frac{\varphi}{\varphi_q^2} - \frac{2n}{\varphi_\text{min}} \left( \frac{\varphi}{\varphi_\text{min}} \right)^{n_1-1} \right)^2, \\
	\label{eq:slow_eta}
	\eta &\simeq -\frac{2}{\varphi_q^2} - \frac{2 n_1(n_1-1)}{\varphi_\text{min}^2} \left( \frac{\varphi}{\varphi_\text{min}} \right)^{n_1-2}, \\
	\zeta &\simeq \left( -\frac{1}{\varphi_l} -2 \frac{\varphi}{\varphi_q^2} - \frac{2n}{\varphi_\text{min}} \left( \frac{\varphi}{\varphi_\text{min}} \right)^{n_1-1} \right) \nonumber \\
	&\qquad \qquad\qquad \left( -2 \frac{2n(n-1)(n-2)}{\varphi_\text{min}^3} \left( \frac{\varphi}{\varphi_\text{min}} \right)^{n_1-3} \right).
\end{align}
To satisfy the slow-roll condition, $\eta$ must be sufficiently smaller than one.
For simplicity, we take $\varphi_q = \sqrt{200}$ in the following.
Note that while $n_s$ and $\dd n_s/\dd\, \text{ln} k$ depend on only $\varphi_l, \varphi_q$ and $\varphi_\text{min}$, 
$A_s$ depends also on $H_\text{first,inf}$ in addition to those parameters.

In Fig.~\ref{fig:allowed_region}, the blue shaded regions show the parameter regions of $\varphi_l$ and $\varphi_\text{min}$ where the corresponding $n_s$ and $\dd n_s/\dd\ln k$ are consistent with the observation.
In each point of this figure, we calculate the value of $H_\text{inf,first}$ with which $A_s$ is consistent with the observation.
The parameter regions which cause the strong resonance are shown by the light black shaded regions. 
One can see that in order for perturbations to avoid the strong resonance 
and for $n_s$ and $\dd \, n_s/\dd\, \text{ln}k$ to be consistent with the observational values,
$H_\text{inf,first}$ must be 
\begin{align}
H_\text{inf,first} \gtrsim
\begin{cases}
\displaystyle
10^{8} \, \text{GeV}
&\text{for}\,\,\,n=3,\\%[1em]
\displaystyle
3.2\times 10^{10} \, \text{GeV}
&\text{for}\,\,\,n=4.\\%[1em]
\end{cases}
\label{eq:H_inf_bound}
\end{align}
If we take $n>4$, the lower bound of $H_\text{inf,first}$ becomes higher 
 because the perturbations in the case of the larger $n$ are accompanied by the stronger resonance.

%%%%%%%%%FIGURE%%%%%%%%%
%%%%%%%%%FIGURE%%%%%%%%%

%%%%%%%%%%%%%%%%%%%%%%%%%%%%
\paragraph*{\bf Concrete example.}
%%%%%%%%%%%%%%%%%%%%%%%%%%%%

%%%%%%%%%FIGURE%%%%%%%%%
%%%%%%%%%FIGURE%%%%%%%%%
%%%%%%%%%FIGURE%%%%%%%%%
\begin{figure}
	\centering
	\includegraphics[width=.40\textwidth]{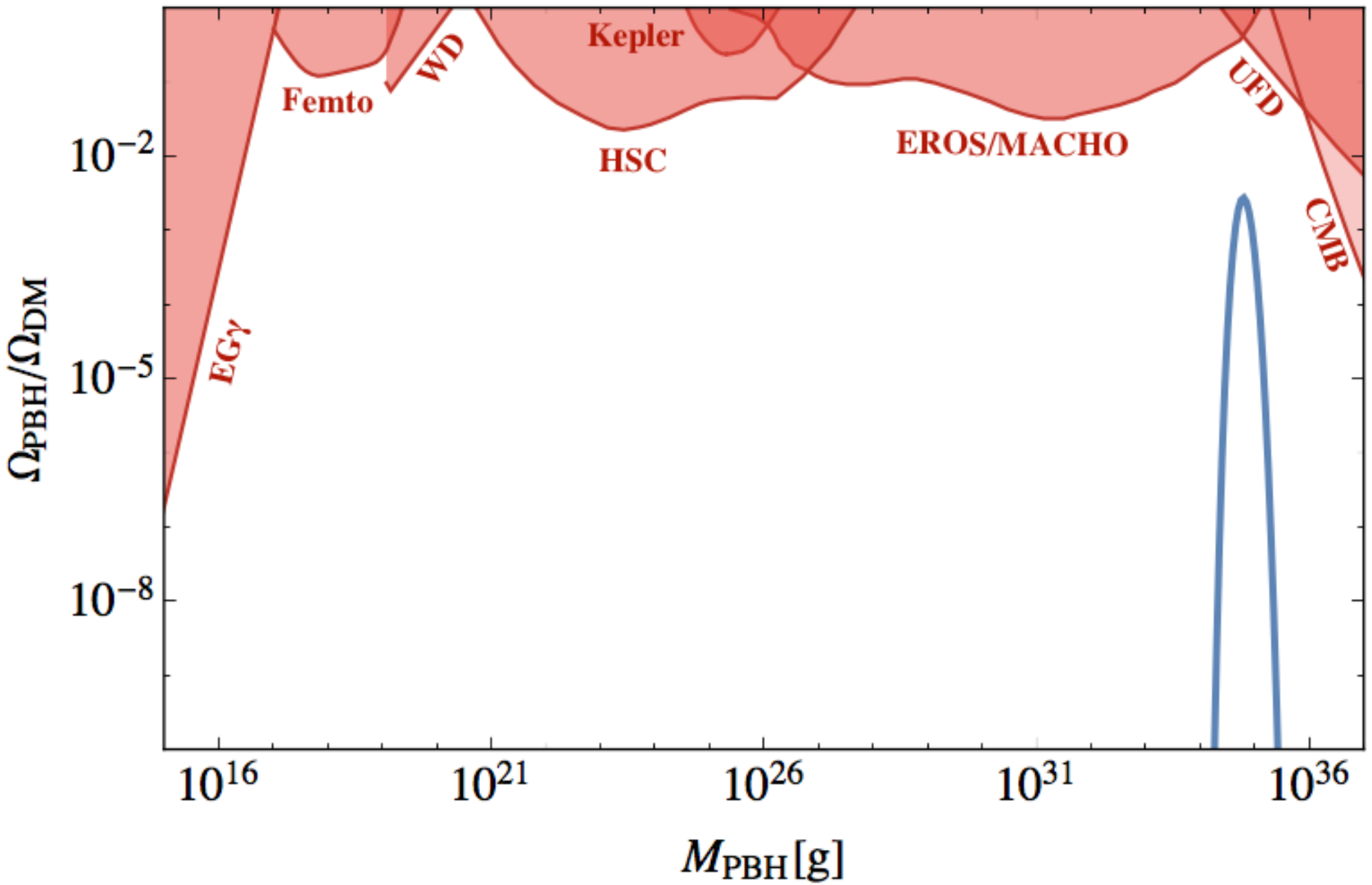}
	\caption{\small
	The PBH mass spectrum for parameters given in Eq.~(\ref{eq:final_para}). 
	Red shaded regions are excluded by 
	extragalactic gamma-rays from Hawking radiation (EG$\gamma$)~\cite{Carr:2009jm}, 
	femtolensing of known gamma-ray bursts (Femto)~\cite{Barnacka:2012bm}, 
	white dwarfs existing in our local galaxy (WD)~\cite{Graham:2015apa}, 
	microlensing search with Subaru Hyper Suprime-Cam (HSC)~\cite{Niikura:2017zjd}, 
	Kepler micro/millilensing (Kepler)~\cite{Griest:2013esa}, 
	EROS/MACHO microlensing (EROS/MACHO)~\cite{Tisserand:2006zx}, 
	dynamical heating of ultra faint dwarf galaxies (UFD)~\cite{Brandt:2016aco}, 
	and accretion constraints from CMB (CMB)~\cite{Ali-Haimoud:2016mbv}.\footnote{
	Recently, Poulin \emph{et al.} have revisited the accretion constraints from CMB taking account of non-spherical accretion 
	and got $\Omega_\text{PBH}/\Omega_\text{DM} \lesssim \mathcal O (10^{-2})$ around $\mathcal O(10) M_\odot$~\cite{Poulin:2017bwe}.
	}
	See also \cite{Carr:2016drx} for a recent summary of observational constraints on PBHs
	and \cite{,Carr:2017jsz,Kuhnel:2017pwq,Green:2016xgy,Inomata:2017okj} for the constraints on the extended PBH mass spectrum.
	}
	\label{fig:pbh_spectrum_n3}
\end{figure}
%%%%%%%%%FIGURE%%%%%%%%%
%%%%%%%%%FIGURE%%%%%%%%%
\begin{figure}
	\centering
	\includegraphics[width=.40\textwidth]{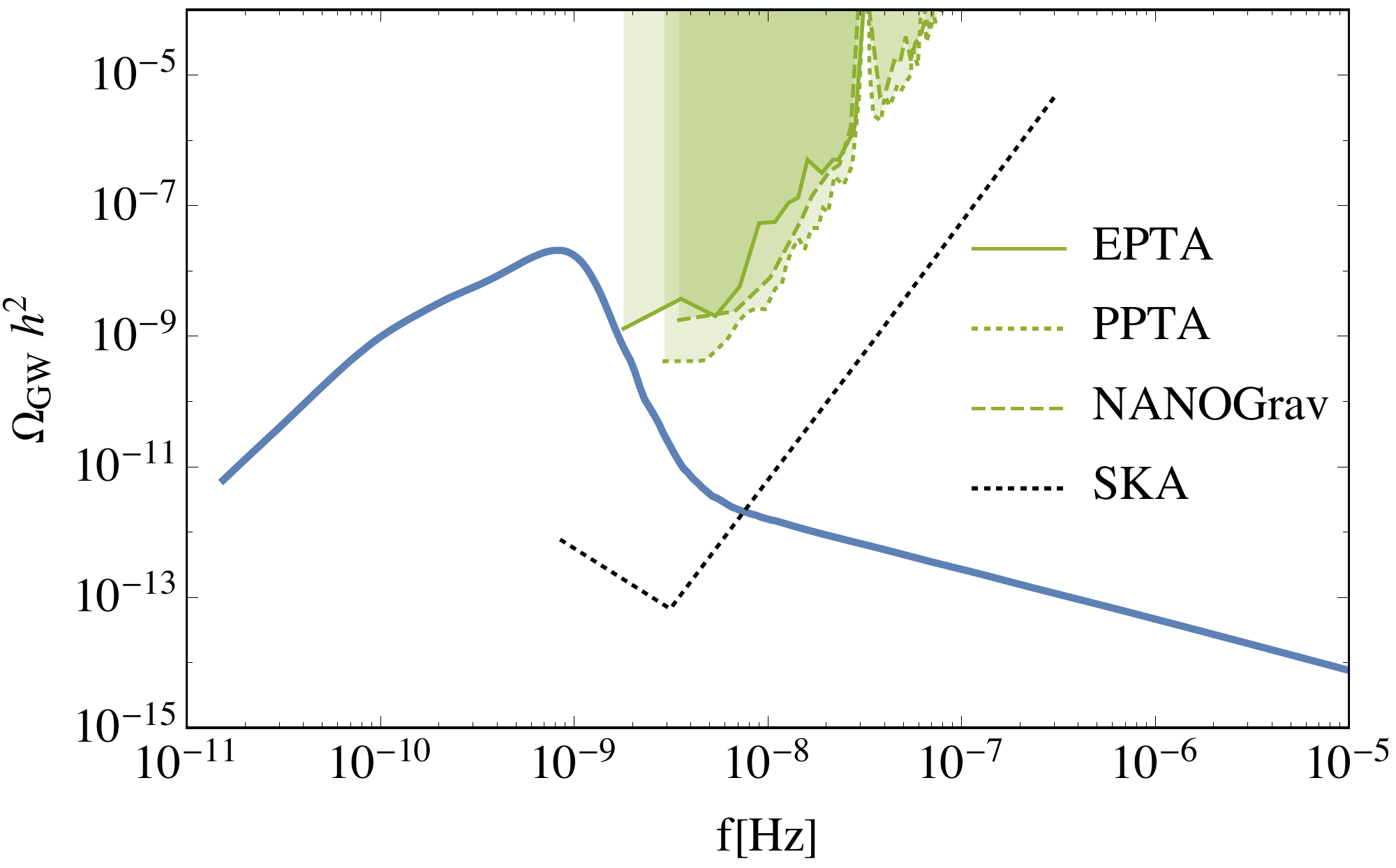}
	\caption{\small
	The induced GW spectrum for parameters given in Eq.~(\ref{eq:final_para}).
	Green shaded regions are excluded by PTA observations with EPTA~\cite{Lentati:2015qwp}, 
	  PPTA~\cite{Shannon:2015ect}, and NANOGrav~\cite{Arzoumanian:2015liz}.
	 A black dotted line shows the prospect of the SKA sensitivity~\cite{Moore:2014lga,Janssen:2014dka}.
	}
	\label{fig:GW_spectrum_n3}
\end{figure}
%%%%%%%%%FIGURE%%%%%%%%%

In this subsection, let us show the full numerical result for the following parameters:
\begin{align}
	\label{eq:final_para}
	&n_1=3, \quad 
	v_1=2.667 \times 10^{-5} \quad(H_\text{inf,first} = 10^9 \, \text{GeV}), \nonumber \\
	\quad &\varphi_l = 7.088\times 10^{5}, 
	\quad \varphi_q= \sqrt{200}, 	
	\quad \varphi_\text{min} = 0.1169, \nonumber \\
	&n_2=3, \quad 	
	v_2 =1.334 \times 10^{-6} \quad(H_\text{inf,second} = 2.5 \times10^6 \, \text{GeV}), \nonumber \\
	\quad &\chi_l = 1.196\times 10^{11}, 
	\quad \chi_q= \sqrt{10}, 	
	\quad \chi_\text{min} = 5.223 \times 10^{-3}, \nonumber \\
	&c_\text{pot}=0.675, \quad 
	c_\text{kin}= -0.671, \quad
	C = 3.465%3
	 \times 10^{-23},
\end{align}
as a fiducial example in the desired parameter regions obtained in the previous subsection.

In Fig.~\ref{fig:curvature_n3}, the resultant linear perturbations are indicated with the constraints by the $\mu$-distortion and BBN.
Though we chose the parameters not to yield the resonance, the power spectrum shows a sharp peak on $k \sim 5\times10^5\,\text{Mpc}^{-1}$.
This is because we apply the amplification mechanism proposed in Ref.~\cite{Inomata:2016rbd}, that is,
$c_\text{pot}$ and $c_\text{kin}$ are chosen so that the effective Hubble-induced mass of 
$\chi$ during the $\varphi$-oscillation phase, $m_\chi^2\simeq\frac{3}{2}(c_\text{pot}+c_\text{kin})H^2$, becomes sufficiently small.
Then the perturbation of $\chi$ can avoid the damping, which leads to the effective amplification of $\delta\chi$.
See  App.~B in Ref.~\cite{Inomata:2016rbd} for a more detailed description. Such a sharp peak is required to avoid the pulsar timing array (PTA) constraints as we will see below.

Using equations described in Sec.~\ref{sec:pbh}, we calculate the expected PBH abundance from the obtained curvature perturbations and 
show its result in Fig.~\ref{fig:pbh_spectrum_n3}.\footnote{
In Fig.~\ref{fig:pbh_spectrum_n3}, we plot the monochromatic observational constraints by the red shaded regions.
Strictly speaking, since our PBH mass spectrum is an extended mass function, the constraints can be tighter as discussed in~\cite{Carr:2017jsz,Kuhnel:2017pwq,Green:2016xgy,Inomata:2017okj}.
To take into account the constraints on the extended mass spectrum,
we have used the analysis described in~\cite{Carr:2017jsz,Inomata:2017okj} and confirmed that our PBH mass spectrum is consistent with the observational constraints.}
From this, one can see that there is a peak at $M_\text{PBH} = 30 M_\odot$ 
and the height is around $\Omega_\text{PBH}/\Omega_\text{DM} \simeq \mathcal O(10^{-3})$.
According to \cite{Sasaki:2016jop}, the expected merger rate in this PBH spectrum is consistent with the merger rate estimated by 
LIGO-Virgo collaboration, $12\text{--}213 \,\text{Gpc}^{-3} \text{yr}^{-1}$~\cite{Abbott:2017vtc}.

As pointed out in Refs.~\cite{Saito:2008jc,Saito:2009jt}, the density (scalar) perturbations large enough to realize abundant PBHs 
can generate GWs (tensor perturbations) due to the second order effect, and the frequency of such secondary GWs for $\calO(10)M_\odot$ PBHs 
lies around the sensitivity region of the PTA experiments $\sim1\,\text{nHz}$. Therefore one has to carefully check the consistency 
with these experiments. Following the instruction of Ref.~\cite{Inomata:2016rbd}, we calculate the current abundance of the secondary GWs 
and show the result in Fig.~\ref{fig:GW_spectrum_n3} with the current and future observational constraints. 
Thanks to the sharpness of our power spectrum of the curvature perturbations,
the resultant GWs successfully evade the current PTA constraints.

%%%%%%%%%%%%%%%%%%%%%%%%%%%%%%%%%%%%%%%%%%%%
\section{Conclusions}
\label{sec:conclusion_a}
%%%%%%%%%%%%%%%%%%%%%%%%%%%%%%%%%%%%%%%%%%%%

In this paper, we have discussed $\mathcal O(10) M_\odot$ PBHs in the presence of string axion DM.

Axion naturally appears in the context of String Theory and can be DM through its coherent oscillation.
Since string axion is light, its perturbations which exit the horizon during inflation behave as isocurvature perturbations constrained by the CMB observations.
The presence of string axion DM puts the severe constraint on the Hubble parameter during the inflation as $H_\text{inf}\lesssim 10^9 \, \text{GeV} \left( f_a / 10^{16}\, \text{GeV} \right)^{0.405}$ 
and therefore the energy scale of inflation must be low.

We have constructed the scenario where the string axion DM and PBHs for the LIGO events coexist, 
taking the double (hilltop + hilltop) inflation model.
We have shown that this inflation model can satisfy the following conditions at the same time.
\begin{itemize}
\item  The energy scale of the inflation model must be low enough to be consistent with the constraint on the isocurvature perturbations. 
\item The large-scale perturbations observed by CMB must be consistent with the observational values of $A_s$, $n_s$, $\dd n_s /\dd \text{ln}k$ and $r_{0.002}$.
\item The perturbations which make $\mathcal O(10) M_\odot$ PBHs must be large enough for the PBH-DM ratio to be $\Omega_\text{PBH}/\Omega_\text{DM} \simeq \mathcal O(10^{-3})$--$\mathcal O(10^{-2})$
and sharp enough to be consistent with the constraints from PTA, $\mu$-distortion, and  BBN observations.
\item The perturbations must remain linear until the second inflation ends to be calculated with the linear analysis.
\end{itemize}
Note that the last condition does not necessarily mean that the non-linear perturbations cannot satisfy the other conditions.
The case with the non-linear perturbations is left for future works.

Finally, let us discuss several outcomes of our scenario.

First, our model predicts the sharp peak around $k\sim 10^{6} $\,Mpc$^{-1}$.
This peak is constrained from the large scale ($\mu$-distortion, BBN) and small scale (PTA).
In other words, our model can be tested by future observations such as PIXIE ($\mu$-distortion) and SKA (PTA).

Second, let us mention the axion miniclusters~\cite{Hogan:1988mp}.
%In the context of axion miniclusters, the scenario where $U(1)_\text{PQ}$ symmetry is broken after inflation has been mainly investigated~\cite{Hogan:1988mp,Kolb:1993zz,Kolb:1993hw}.
It has been suggested that axion can form a dense clump by e.g. large isocurvature perturbations associated with the breakdown of $U(1)_\text{PQ}$ symmetry after inflation~\cite{Hogan:1988mp,Kolb:1993zz,Kolb:1993hw}.
Although we have considered the scenario where $U(1)_\text{PQ}$ symmetry is broken before inflation, our scenario expects the large adiabatic perturbations around $k\sim 10^{6}$\,Mpc$^{-1}$
and hence the axion miniclusters might be produced.\footnote{
%Dense axion clumps produced by the primordial perturbations are remarked in Ref.~\cite{Tkachev:1991ka}.
Tkachev has remarked the possibility that the dense axion clumps are produced by the large primordial perturbations in Ref.~\cite{Tkachev:1991ka}.
}
%This situation has been remarked by Tkachev in the context of the bose-star formation~\cite{Tkachev:1991ka}.
Since the perturbations are smaller than that in the scenario where $U(1)_\text{PQ}$ is broken after inflation (but still large),
we can expect that the produced axion miniclusters in our scenario are sparser~\cite{Kolb:1994fi}.
As discussed in~\cite{Tinyakov:2015cgg},
the sparse axion miniclusters tend to get disrupted by the encounters with the stars in the Galactic disk and the fragments of the destructed miniclusters form tidal streams.
Although there are still many uncertainties about this phenomenon,
it is possible that the tidal streams can be detected in the future.

%%%%%%%%%%%%%%%%%%%%%%%%%%%%%%%%%%
%%%%%%%%%%% Acknowledge %%%%%%%%%%%
%%%%%%%%%%%%%%%%%%%%%%%%%%%%%%%%%%
\section*{Acknowledgements}
{\small
\noindent
This work is supported by Grant-in-Aid for Scientific Research from the Ministry of Education,
Science, Sports, and Culture (MEXT), Japan,  No.\ 15H05889 (M.K.), No.\ 17K05434 (M.K.), No.\ 17H01131 (M.K.), %25400248 (M.K.),
No.\ 26104009 (T.T.Y.), No.\ 26287039 (T.T.Y.) and No.\ 16H02176 (T.T.Y.), 
World Premier International Research Center Initiative (WPI Initiative), MEXT, Japan 
(K.I., M.K., K.M., Y.T., and T.T.Y.),
JSPS Research Fellowships for Young Scientists (Y.T., and K.M.),
R\'egion \^Ile-de-France (Y.T.),
and Advanced Leading Graduate Course for Photon Science (K.I.).
}

\appendix
%%%%%%%%%%%%%%%%%%%%%%%%%%%%%%%%%%
%%%%%%%%%%% Appendix %%%%%%%%%%%
%%%%%%%%%%%%%%%%%%%%%%%%%%%%%%%%%%

 %%%%%%%%%%%%%%%%%%%%%%%%%%%%
\section{Resonance}
\label{sec:resonance}
 %%%%%%%%%%%%%%%%%%%%%%%%%%%%
 
In this appendix, we discuss the resonances which occur after hilltop inflation.

For simplicity,
we consider the toy inflation model with the following potential:
\begin{align}
	V(\phi)
		&=v^4\left( 1- \frac{\phi^n}{\phi_\text{min}^n} \right)^2.
	\label{eq:ex_inf_pot}
\end{align}
If we take $v$ and $\phi_\text{min}$ as $v_1$ and $\varphi_\text{min}$, 
this potential is a good approximation of Eq.~(\ref{eq:inf_potential_new1}) during the oscillation of $\varphi$ which follows the first inflation.
In this inflation model, after inflation, $\phi$ starts to oscillate around $\phi_\text{min}$.
We define $\phi_e$ as the value at the end of inflation and $\phi_m$ as the value which satisfies $V''(\phi_m) =0$,
where the prime denotes a derivative with respect to $\phi$, e.g. $V''(\phi) = \partial^2 V(\phi)/\partial \phi^2$.
%%%%%%%%%FIGURE%%%%%%%%%
\begin{figure}
	\centering
	\includegraphics[width=.50\textwidth]{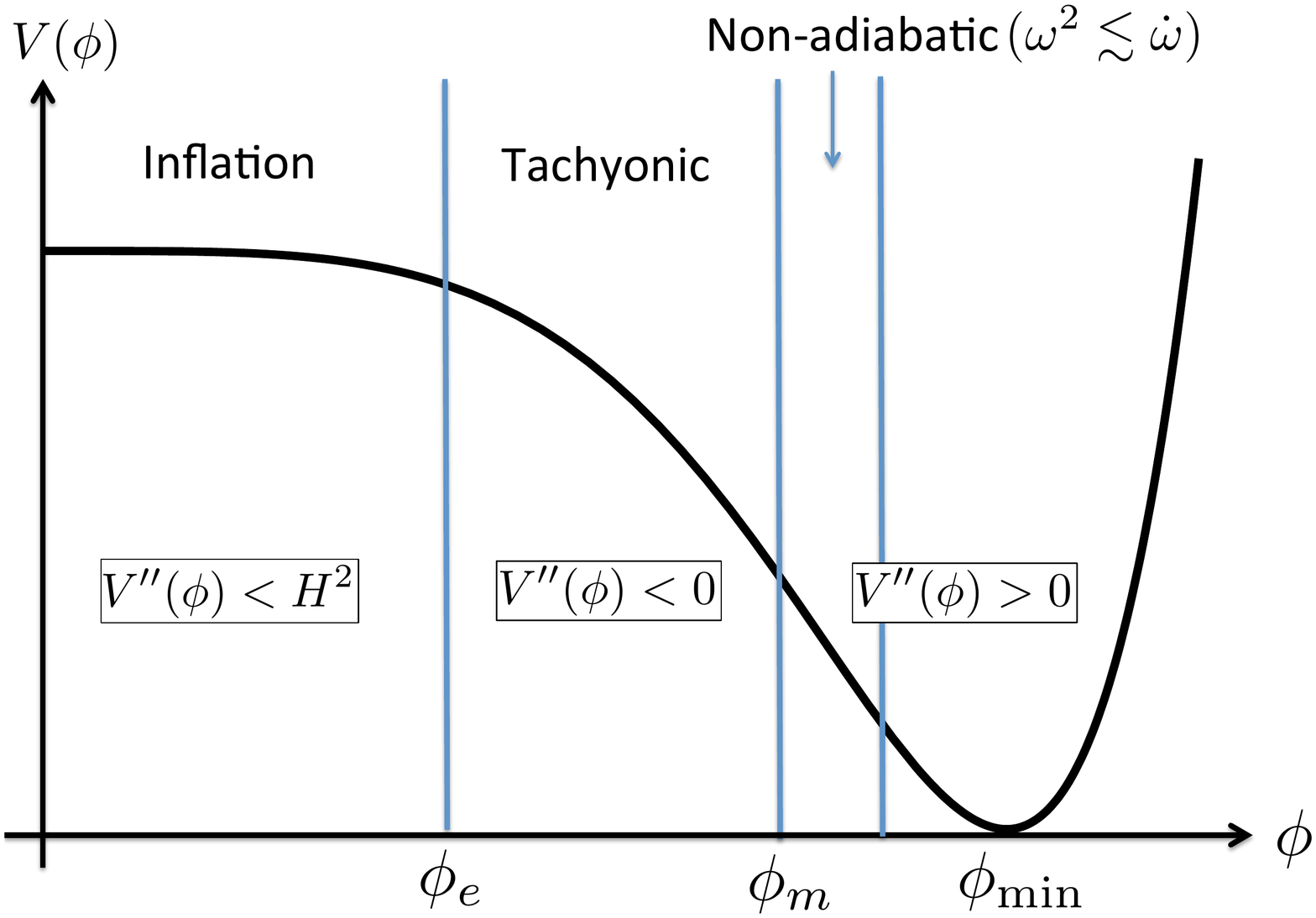}
	\caption{\small
	The schematic image of the inflation potential given by Eq.~(\ref{eq:ex_inf_pot}).
	This figure is a modified version of Fig. 1 in \cite{Brax:2010ai}.
	}
	\label{fig:new_inf_pot}
\end{figure}
%%%%%%%%%FIGURE%%%%%%%%%
Fig. \ref{fig:new_inf_pot} shows the schematic image of the hilltop inflation potential.
We can derive the value of $\phi_e$ and $\phi_m$ as 
\begin{align}
	\label{eq:phi_e}
	\phi_e = \left( \frac{\phi_\text{min}^n}{2n(n-1)} \right)^{\frac{1}{n-2}}, \\
	\label{eq:phi_m}
	\phi_m = \left( \frac{n-1}{2n-1} \right)^{\frac{1}{n}} \phi_\text{min},
\end{align}
where we have used the equality $|V''(\phi_e)/V| = 1$ to derive $\phi_e$.

This toy inflation model goes through the following steps.
When $\phi<\phi_e$, $\phi$ slowly rolls down its potential and inflation occurs with $H_\text{inf} = v^2/\sqrt{3}$.
After $\phi$ passes through $\phi_e$, the slow-roll inflation ends and $\phi$ starts to oscillate around $\phi_\text{min}$.
At first, the amplitude of oscillations is large enough so that $\phi$ can return to the tachyonic region, $\phi<\phi_m$.
During this period, the perturbations of $\phi$ grow rapidly due to the tachyonic resonance.
Later, the amplitude becomes small due to the expansion of the Universe and $\phi$ cannot return to the tachyonic region.
Although the tachyonic resonance stops, perturbations still grow due to the non-adiabatic resonance.
Adiabatic condition, $\omega^2 \gg \dot \omega$ ($\omega^2 \equiv V''(\phi)$), fails violently when $\phi \sim \phi_m$ because $V''(\phi)\sim 0$ at $\phi \sim \phi_m$.
Finally, the amplitude of $\phi$ becomes so small that $\phi$ does not pass through the non-adiabatic region and the perturbations decay due to Hubble friction.

%%%%%%%%%FIGURE%%%%%%%%%
\begin{figure}
	\centering
	\includegraphics[width=.40\textwidth]{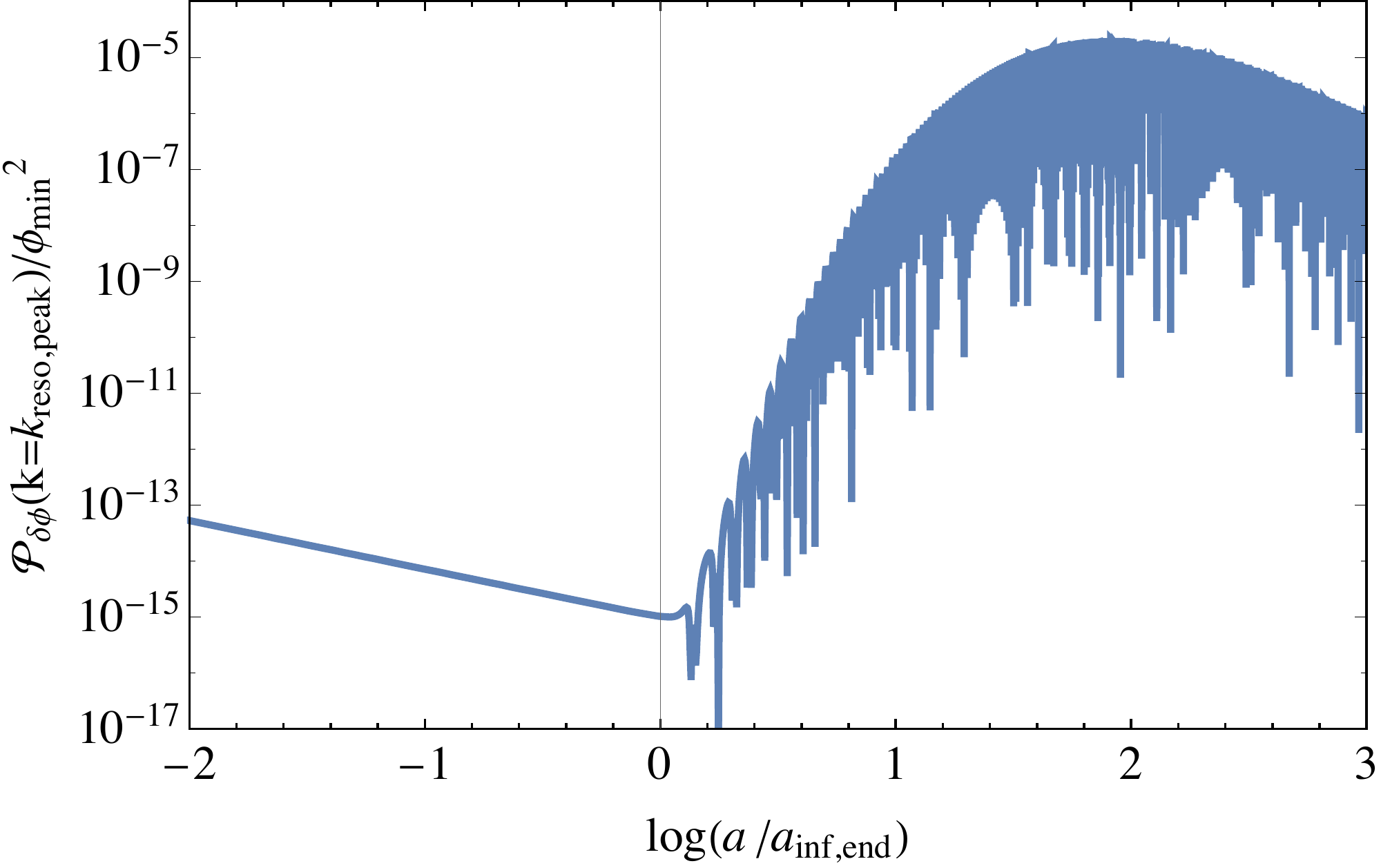}
	\caption{\small
	The time evolution of $\mathcal P_{\delta \phi}(k=k_\text{peak})/\phi_\text{min}^2$ with $n=4, H_\text{inf}=10^9 \,\text{GeV}$, and $\phi_\text{min}=0.109$.
	We use the scale factor as time variables, where $a_\text{inf,end}$ is the scale factor at the inflation end.
	}
	\label{fig:reso_evol}
\end{figure}
%%%%%%%%%FIGURE%%%%%%%%%

Fig. \ref{fig:reso_evol} shows the time evolution of power spectrum of $\delta \phi$ divided by $\phi_\text{min}^2$ at the resonance peak scale, 
$\mathcal P_{\delta \phi}(k=k_\text{reso,peak})/\phi_\text{min}^2$, with $n=4, \,H_\text{inf}=10^9\,\text{GeV}$, and $\phi_\text{min}=0.109$.
$\mathcal P_{\delta \phi}(k=k_\text{reso,peak})/\phi_\text{min}^2$ roughly corresponds to the squared ratio of the perturbation to the background field value.
If $\mathcal P_{\delta \phi}(k=k_\text{reso,peak})/\phi_\text{min}^2 > 1$, we regard the perturbation as non-linear.
From Fig. \ref{fig:reso_evol}, we can see that the perturbation grows until $\text{log}(a/a_\text{inf,end})\sim2$ 
due to the tachyonic and non-adiabatic resonance and after the growth ($\text{log}(a/a_\text{inf,end})\geq2$), 
the perturbation decays due to Hubble friction. 
Here, we define the conformal time when a perturbation becomes a maximum as $\eta_\text{peak}$, 
which corresponds to the conformal time at $\text{log}(a/a_\text{inf,end}) \sim 2$ in Fig.~\ref{fig:reso_evol}.

Tachyonic resonance has been already discussed analytically in~\cite{Brax:2010ai}.
Following Ref.~\cite{Brax:2010ai}, we can estimate the peak scale of the tachyonic resonance as
\begin{align}
	\frac{k_\text{reso,peak}}{a_\text{inf,end}} 
	\simeq \left( \phi_\text{min} \right)^{\frac{n-2}{2n}} m_\phi = \sqrt{2} n v^2 \left( \phi_\text{min} \right)^{-\frac{n+2}{2n}},
	\label{eq:peak_loc_app}
\end{align}
where $k_\text{reso,peak}$ is the comoving wave number at the peak, $m_\phi$ is the mass of $\phi$ around $\phi_\text{min}$ defined as $m_\phi = \sqrt{2} n v^2/\phi_\text{min}$
and $a_\text{inf,end}$ is the scale factor at the inflation end.
Since the resonances occur so rapidly and the peak scale is determined at the beginning of the resonances, $a_\text{inf,end}$ appears in Eq.~(\ref{eq:peak_loc_app}).
We confirm that our numerical results are consistent with Eq.~(\ref{eq:peak_loc_app}).
Note that the resonance peak scale is different from the peak scale of the curvature perturbation which produces PBHs in our model ($k_\text{reso,peak} \neq k_\text{peak}$).

\begin{figure}[htbp]
	\centering
	\vspace{10pt}
	\includegraphics[width=.45\textwidth]{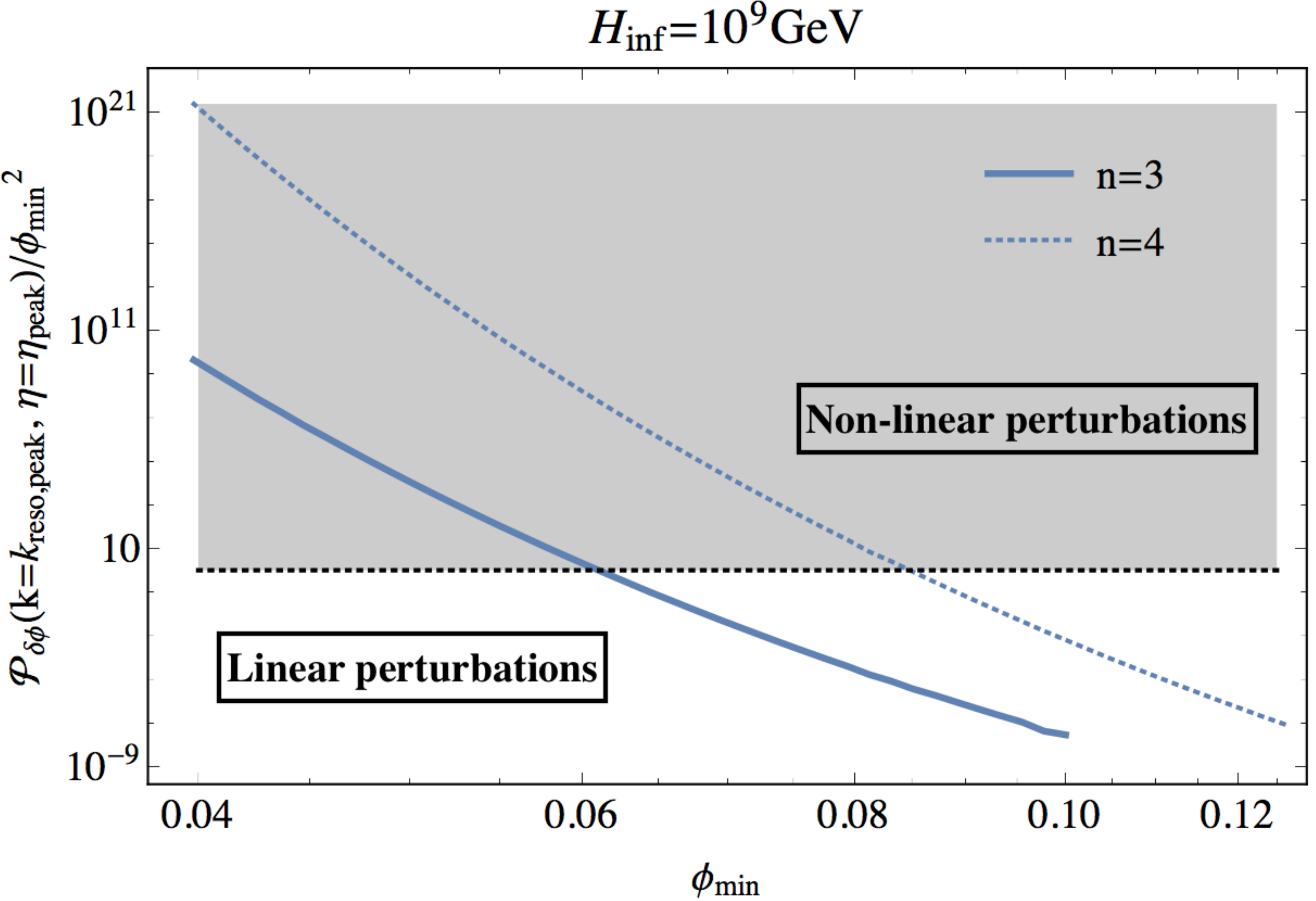}
	\caption{\small
	The $\phi_\text{min}$ dependence of $\mathcal P_{\delta \phi} (k=k_\text{reso,peak}, \eta = \eta_\text{peak})/ \phi^2_\text{min}$.	
	We take $H_\text{inf} = 10^9 \,\text{GeV}$.
	The blue solid (dotted) line shows $\mathcal P_{\delta \phi} (k=k_\text{reso,peak}, \eta = \eta_\text{peak})/ \phi^2_\text{min}$ in the case of $n=3$ ($n=4$).
	The gray shaded region shows $\mathcal P_{\delta \phi} (k=k_\text{reso,peak}, \eta = \eta_\text{peak})/ \phi^2_\text{min}>1$, where perturbations become non-linear.
	}
	\label{fig:phimin_vs_peak}
\end{figure}

Fig.~\ref{fig:phimin_vs_peak} shows the  $\phi_\text{min}$ dependence of $\mathcal P_{\delta \phi} (k=k_\text{reso,peak}, \eta = \eta_\text{peak})/ \phi^2_\text{min}$.
From this figure, we can see that the decrease of $\phi_\text{min}$ or the increase of $n$ leads to the increase of the peak height.
This results can be understood as follows.
The peak height depends on the relation between  the timescale of the oscillation and that of Hubble friction.
The timescale of the oscillation corresponds to $m_\phi^{-1} = \phi_\text{min}/(\sqrt{2} n v^2)$.
On the other hand, the timescale of Hubble friction is related to the Hubble parameter as $H_\text{inf}^{-1} \simeq 1/(\sqrt{3} v^2)$.
The important factor is the ratio of $m_\phi^{-1}$ to $H_\text{inf}^{-1}$.
If we take a smaller value of $\phi_\text{min}$ or a larger value of $n$, only $m_\phi^{-1}$ becomes smaller (but $H_\text{inf}^{-1}$ does not change).
Then the inflaton can pass through the resonance region more times 
before its amplitude becomes small due to Hubble friction and the peak height becomes larger.

Finally, let us mention the $H_\text{inf}$ dependence of the peak height.
To understand it analytically, we neglect the metric perturbations\footnote{
We have numerically checked that the metric perturbations have little effect on the dynamics of $\delta \phi$ during the resonances.
} and then the equation of motion for $\delta \phi$ which corresponds to the peak scale can be written as
\begin{align}
	\allowdisplaybreaks[4]
	&\ddot \delta \phi + 3H \dot \delta \phi + \frac{k_\text{reso,peak}^2}{a_\text{inf,end}^2} \delta \phi + V''(\phi) \delta \phi = 0 \nonumber \\
	\Rightarrow\ 
	&\ddot \delta \phi + 3H \dot \delta \phi  + 2n^2 v^4 \left( \phi_\text{min} \right)^{-\frac{n+2}{n}} \delta \phi \nonumber \\
	& \qquad \qquad \qquad \qquad 
	+ v^4 \frac{\partial^2}{\partial \phi^2} \left( 1 - \left( \frac{\phi}{\phi_\text{min}} \right)^n \right)^2 \delta \phi = 0 \nonumber \\
	\Rightarrow\ 	
	&\frac{\dd^2}{\dd \tau^2} \delta \phi + 3\tilde{H} \frac{\dd}{\dd \tau} \delta \phi  + 2n^2 \left( \phi_\text{min} \right)^{-\frac{n+2}{n}} \delta \phi \nonumber \\
	& \qquad \qquad \qquad \qquad 
	+ \frac{\partial^2}{\partial \phi^2} \left( 1 - \left( \frac{\phi}{\phi_\text{min}} \right)^n \right)^2 \delta \phi = 0,
	\label{eq:eom_perturbations}
\end{align}
where $\dd / \dd \tau \equiv \dd / \dd (v^2 t)$ and $\tilde H \equiv \frac{1}{a} \frac{\dd a}{\dd \tau}$.
From Eq.~(\ref{eq:eom_perturbations}), we can see that $H_\text{inf} (=v^2/\sqrt{3})$ determines only the timescale of the perturbation evolution.
Note that even though the growth rate of perturbations due to the resonances are independent of $H_\text{inf}$, 
the value of $\delta \phi$ before the resonances depends on $H_\text{inf}$ as $\delta \phi \simeq H_\text{inf}/2\pi$.
Therefore, we can estimate the $H_\text{inf}$ dependence as $\mathcal P_{\delta \phi} (k=k_\text{reso,peak}, \eta = \eta_\text{peak})/ \phi^2_\text{min} \propto H_\text{inf}^2$.
This estimation is consistent with our numerical calculation which includes the metric perturbations.
Using the result of Fig.~\ref{fig:phimin_vs_peak} and this assumption about $H_\text{inf}$-dependence, we illustrate the model parameters with which
the perturbations become non-linear by the light black shaded regions in Fig.~\ref{fig:allowed_region}. We avoid these regions for a concrete example
in this paper.

%%%%%%%%%FIGURE%%%%%%%%%
%%%%%%%%%FIGURE%%%%%%%%%

%%%%%%%%%%%%%%%%%%%%%%%%

%\section{Analytic comprehension for multiple horizon crossing modes}
\section{Relation between the peak scale and the first inflation end scale}
\label{sec:multihorizon}

In this appendix, we derive Eq.~(\ref{eq:end_peak_e_fold}).
We define $a_\text{peak}$ as the scale factor which satisfies $a_\text{peak} H_\text{inf,first} = k_\text{peak}$, 
$a_\text{first,e}$ as the scale factor at the first inflation end and $a_\text{second,b}$ as the scale factor at the second inflation beginning.
The scale factor dependence of $aH$ is given by
\begin{align}
aH \propto
\begin{cases}
\displaystyle
a
&\text{during the first inflation},\\%[1em]
\displaystyle
a^{-1/2}
&\text{during the oscillation of inflaton}\,\  \varphi.%[1em]
\end{cases}
\label{eq:aH_behavior}
\end{align}
Taking account of the fact that $H$ is proportional to $a^{-3/2}$ during the oscillation phase,
we get the following equation:
\begin{align}
 \text{log} \left( \frac{H_\text{inf,first}}{H_\text{inf,second}} \right)
= \frac{3}{2} \text{log} \left( \frac{a_\text{second,b}}{a_\text{first,e}}\right).
\label{eq:h_a_rel}
\end{align}
Following the discussion of App.~B in Ref.~\cite{Inomata:2016rbd}, we can estimate as $k_\text{peak} = a_\text{second,b} H_\text{inf,second} \,(= a_\text{peak} H_\text{inf,first} )$.
Using this estimation and Eq.~(\ref{eq:aH_behavior}), we get the following equation:
\begin{align}
\text{log} \left( \frac{a_\text{second,b}}{a_\text{first,e}}\right)
&= 2 \text{log} \left( \frac{a_\text{first,e} H_\text{inf,first}}{a_\text{second,b} H_\text{inf,second} }\right) \nonumber \\
&= 2 \text{log} \left( \frac{a_\text{first,e}}{a_\text{peak}} \right).
\label{eq:h_a_rel_peak}
\end{align}
From Eqs.~(\ref{eq:h_a_rel}) and~(\ref{eq:h_a_rel_peak}), we can derive Eq.~(\ref{eq:end_peak_e_fold}) as
\begin{align}
	\text{log} \left( \frac{H_\text{inf,first}}{H_\text{inf,second}} \right)
	&= 3 \text{log} \left( \frac{a_\text{first,e}}{a_\text{peak}}\right) \\
	\Rightarrow \ \ 
	\frac{H_\text{inf,first}}{H_\text{inf,second}} &= \ee^{3(N_\text{peak} - N_\text{first inf, end}) }.
	\label{eq:end_peak_e_fold_test}
\end{align}
%%

%%%%%%%%%%%%%%%%%%%%%%%%%%%%%%%%%
%%%%%%%%%%% References %%%%%%%%%%%
%%%%%%%%%%%%%%%%%%%%%%%%%%%%%%%%%
\small
\bibliographystyle{apsrev4-1}

\end{document}